\documentclass[preprint,12pt]{elsarticle}
\usepackage{graphicx,amscd,amsmath,amssymb,amsfonts,verbatim,bm, dsfont}
\usepackage{float}
\usepackage{tikz}
\usepackage{placeins}
\usepackage{booktabs}
\usepackage{colortbl}
\usepackage{url}
\usepackage{threeparttable}
\usetikzlibrary{spy}
\usetikzlibrary{backgrounds}
\usetikzlibrary{shapes,decorations,arrows,calc,arrows.meta,fit,positioning}
\usepackage{subcaption}

\usepackage{amssymb}

\journal{Computational Statistics and Data Analysis}

\begin{document}

\begin{frontmatter}



\title{Plasmode simulation for the evaluation of
causal inference methods in homophilous social networks}


\author[inst1]{Vanessa McNealis}

\affiliation[inst1]{organization={Department of Epidemiology and Biostatistics, McGill University},
            addressline={2001 McGill College, Suite 1200}, 
            city={Montreal},
            postcode={H3A 1G1}, 
            state={Quebec},
            country={Canada}}

\author[inst1]{Erica E. M. Moodie}
\author[inst2]{Nema Dean}

\affiliation[inst2]{organization={School of Mathematics and Statistics, University of Glasgow},
            addressline={University Place}, 
            city={Glasgow},
            postcode={G12 8QW}, 
            state={Scotland},
            country={United Kingdom}}

\begin{abstract}
Typical simulation approaches for evaluating the performance of statistical methods on populations embedded in social networks may fail to capture important features of real-world networks. It can therefore be unclear whether inference methods for causal effects due to interference that have been shown to perform well in such synthetic networks are applicable to social networks which arise in the real world. Plasmode simulation studies use a real dataset created from natural processes, but with part of the data-generation mechanism known. However, given the sensitivity of relational data, many network data are protected from unauthorized access or disclosure. In such case, plasmode simulations cannot use released versions of real datasets which often omit the network links, and instead can only rely on parameters estimated from them. A statistical framework for creating replicated simulation datasets from private social network data is developed and validated. The approach consists of simulating from a parametric exponential family random graph model fitted to the network data and resampling from the observed exposure and covariate distributions to preserve the associations among these variables.
\end{abstract}

\begin{keyword}
simulation \sep plasmode \sep social network \sep causal inference \sep interference \sep spillover \sep homophily
\end{keyword}

\end{frontmatter}


\section{Introduction}
\label{sec:introduction}

In recent years there has been an increasing interest in evaluating causal effects in the presence of interference (also known as dissemination or spillover), which arises when the treatment assigned to an individual influences not only their own outcome directly but also those of their contacts. To this end, social network data are particularly useful and rich, since they provide information about connections between individuals and thus the possible pathways to interference. However, causal inferences on social network data may suffer from bias due to network confounding, and it is unclear whether existing methods for confounding adjustment that have been shown to perform well in small, synthetic observational social networks are applicable to real-world networks. Homophily, which is defined as the increased tendency of units with similar characteristics of forming ties, constitutes one source of network confounding. Homophily confounding has been shown to threaten the inference for causal effects, and existing inference methods have predominantly been evaluated in computer-generated networks or real networks that are devoid of homophily confounding \citep{lee2021estimating, forastiere2021identification, liu2023nonparametric, mcnealis2023doubly}.

In the context of statistical methodology research, typical Monte Carlo simulation approaches often use artificial data-generating processes which may be adapted such that a method favored by a researcher appears superior, and can thus lead to overoptimism in comparative performance assessments \citep{pawel2024pitfalls, buchka2021optimistic}. Further, simulated data may fail to capture important features of real-world datasets. This is particularly the case for social network data. For instance, observational network studies can comprise dozens of measured nodal attributes with complex covariance structures. These covariates may also be associated with the network structure itself, as they can influence the propensity of a tie forming between two nodes (i.e., a social link forming between two individuals). Such complex dependencies are rarely adequately captured in simulated network data. 



A popular remedy is to augment real data with simulated data to create a \emph{plasmode}, that is, a real dataset created from natural processes, but with part of the data-generation mechanism known (e.g., generation of synthetic exposure and outcome variables) \citep{vaughan2009use, franklin2014plasmode}. Studies of methods' performance that leverage plasmodes are termed \emph{plasmode simulations}. However, when access to the real data of interest is restricted and limited to a physical or virtual enclave, it can be complicated or even infeasible to perform a plasmode simulation. \emph{Synthetic} plasmode simulations utilize fully simulated data, but induce dependence between variables so as to match the empirical joint distribution in a real dataset. Extending this idea to relational data, it would be conceivable to simulate social networks from a given statistical network model, whose parameters have been estimated beforehand on the private network. Provided that the release of sufficient statistics is authorized by the data use agreement, it is always possible for a researcher to carry out a synthetic plasmode simulation study in their own computing enviroment.  

In this paper, we outline a statistical and computational framework for creating replicated
simulation datasets based on an empirical cross-sectional network. Notably, we describe a method for sharing social network data including the associated nodal attributes for the purpose of simulation, which can be used when faced with restricted-use relational data. The objective of this work is to enable the evaluation of approaches to spillover effect estimation in simulated networks that preserve the complex features of social network data but also have a known true interference effect. To assess and illustrate the approach, we consider data from the National Longitudinal Study of Adolescent Health.

The paper is structured as follows. In Section \ref{section:preliminaries}, we describe foundational concepts needed for causal inference with network interference and exponential random graph models. In Section \ref{section:methods}, we describe the synthetic plasmode simulation framework. In Section \ref{section:application}, we motivate our framework with an empirical investigation of causal inference methods for social network data in the presence of unaddressed homophily. We close with a general discussion in Section \ref{section:discussion}.

\section{Preliminaries}
\label{section:preliminaries}
 In this work, we restrict ourselves to point-treatment or cross-sectional studies, where the social network is assumed to be static from the time of recruitment to the time of outcome assessment. We consider a binary undirected network $G=(\mathcal{N}, \mathbb{E})$, where $\mathcal{N} = \{1, 2, \ldots, N\}$ is a set of $N$ nodes or vertices and $\mathbb{E}$ is a set of edges or ties with generic element $\{i,j\} = \{j,i\}$ denoting the presence of an edge between nodes $i$ and $j$. Any pair of vertices $i$ and $j$, $i \neq j$, is referred to as a dyad. 
For convenience, we represent the network $G$ by the sociomatrix $\bm{A}$, which has entries $A_{ij} = 1$ if $\{i,j\} \in \mathbb{E}$ and 0 otherwise. Since we assume that there are no self-loops, the diagonal elements of $\bm{A}$ are identically $0$. Because $\bm{A}$ is symmetric, the network consists of $\binom{N}{2} = N(N-1)/2$ unordered dyads $\{i,j\}$. We define the neighborhood $\mathcal{N}_i$ of node $i$, $i=1,2,\ldots, N$, as the set of nodes $j$ which satisfy $A_{ij} = 1$ and the cardinality of $\mathcal{N}_i$ is denoted as $|\mathcal{N}_i| = d_i$. Additionally, we denote $\mathcal{N}_{i}^* = \mathcal{N}_{i} \cup {i}$ as the union of the node $i$ and the nodes in $\mathcal{N}_i$.   

   For each node $i$, $i=1,2,\ldots, N$, we observe the node-level vector of attributes $O_i = (\bm{X}_i, Z_i, Y_i)$, where $\bm{X}_i=(X_{i1}, \ldots, X_{ip})'$ denotes a $p$-dimensional vector of pretreatment covariates, $Z_i$ denotes a binary (possibly self-selected) treatment with $Z_i = 1$ if treated and $0$ otherwise, and $Y_i$ is a univariate, continuous outcome of interest. We assume that covariates precede the treatment or exposure assignment and the formation of network ties. We also assume that the outcome is assessed after the assignment of the treatment or exposure. We let $\bm{X}_Y$ and $\bm{X}_G$ denote the sets of pretreatment covariates related to the outcome generating mechanism and the formation of ties in $G$, each of dimension $p_1$ and $p_2$, respectively. 
 Further, let $\bm{X}_{\mathcal{N}_i} = (\bm{X}_{j_1}, \ldots, \bm{X}_{j_{d_i}})'$ and $\bm{Z}_{\mathcal{N}_i} = (Z_{j_1}, \ldots, Z_{j_{d_i}})'$ denote the neighborhood matrix of pretreatment covariates and neighborhood vector of treatments for node $i$, respectively, where $j_{k} \in \mathcal{N}_i, k = 1, \ldots, d_i$.

In what follows, private data structures refer to data structures that are sensitive in nature and made available to researchers through highly controlled data access mechanisms. We always assume that the relational data encoded in $G$ is private and describe ways to release the information contained in private nodal attributes. Public data structures taken to be any structures that do not satisfy our definition of private data structures.

\subsection{Causal inference on social networks and inferential targets}

For the sake of simplicity and conciseness, let us assume that the linear structural equation model that explains the outcome of node $i$, for $i = 1, \ldots, N$, is given by
\begin{equation} 
\label{eqn:modely}
Y_{i} = \beta_0 + \beta_Z Z_{ i} + \beta_{\bm{Z}_{\mathcal{N}}} \frac{\sum_{j \in \mathcal{N}_i} Z_j}{d_i} + \beta_{Z \times \bm{Z}_{\mathcal{N}}}Z_{ i} \frac{\sum_{j \in \mathcal{N}_i} Z_j}{d_i}+  \bm{X}_{Y i} ^{\top} \bm{\beta}_{\bm{X}_Y} + \varepsilon_{i},
\end{equation}
 where $\bm{X}_Y$ represents the set of nodal covariates that predict the outcome and $\varepsilon_i \sim N(0, \sigma^2_{\varepsilon})$ is measurement error. This model states that the outcome of node $i$ is affected not only by their own treatment $Z_i$ but also by the average treatment in the neighborhood, $\frac{\sum_{j \in \mathcal{N}_i} Z_j}{d_i}$, which describes a form of interference. Previous authors have coined this assumption as \textit{neighborhood interference}, since the outcome of node $i$ only depends on the treatments received by their close neighbors, but not on those received by higher-order neighbors \citep{forastiere2021identification, lee2021estimating}. Note that the proportion of treated neighbors for node $i$ can also be expressed as $\frac{\sum_{j=1}^N A_{ij} Z_j}{\sum_{j=1}^n A_{ij}}$, which highlights how the edge and nodal attribute data are tied in this structural equation model. 

We use the potential outcomes approach to causal inference and denote a potential outcome for node $i$ under individual treatment $z_i$ and neighborhood treatment $\sum_{j \in \mathcal{N}_i} z_j$ as $y_i(z_i, \sum_{j \in \mathcal{N}_i} z_j)$.  
 While several estimands have been proposed for this setting, we focus on those based on the Bernoulli allocation framework, which standardizes individual potential outcomes with respect to a counterfactual population in which treatment is administered to each individual independently with probability $\alpha$ \citep{tchetgen2012causal, papadogeorgou2019causal, lee2021estimating, mcnealis2023doubly}. We define the $i$-th individual's average potential outcome conditional on individual exposure $z$ and allocation strategy $\alpha$ as \begin{equation}
 \label{eqn:indapo}
 \bar{y}_{i}(z;\alpha) = \sum_{\Sigma \bm{z}_{\mathcal{N}_i} = 0 }^{d_i} y_{i}(z, \Sigma \bm{z}_{\mathcal{N}_i}) \pi( \Sigma \bm{z}_{\mathcal{N}_i}; \alpha).    
 \end{equation}
We can define meaningful causal contrasts from the individual average potential outcome. The individual direct effect of exposure under allocation strategy $\alpha$ is defined as \begin{equation}
\label{eqn:de}
DE_{i}(\alpha) = \bar{y}_{i}(1;\alpha) - \bar{y}_{i}(0;\alpha).
\end{equation}
The individual indirect effect of exposure, which compares distinct allocation strategies $\alpha$ and $\alpha'$ among the unexposed, is defined as
\begin{equation}
\label{eqn:ie}
    IE_{i}(\alpha, \alpha') = \bar{y}_{i}(0;\alpha) - \bar{y}_{i}(0;\alpha').
\end{equation}
While the individual effects are not identifiable, inference about population averages of these quantities can be carried out from social network data.  
Conditions to identify and estimate average causal effects in the presence of observational network interference are well documented \cite{forastiere2021identification, lee2021estimating, mcfowland2023estimating, mcnealis2023doubly}. Assumptions include extensions of the usual causal assumptions including consistency and conditional exchangeability to the network interference setting \citep{liu2016inverse, forastiere2021identification, lee2021estimating, mcnealis2023doubly}. For instance, under causal consistency, the equality $Y_i = y_i(Z_i, \sum_{j \in \mathcal{N}_i} Z_j)$ holds. If the causal assumptions hold, then meaningful causal estimands can be identified from Model (\ref{eqn:modely}).

\subsection{Exponential random graph models}
\label{section:prelim_ERGM}
A common approach to generating the social network of the population is by simulating from random graph models. We pursue this within the framework of exponential random graph models (ERGMs), a class of generative models for modeling network dependence based on exponential family distribution theory \citep{robins2007introduction, hunter2008ergm}. ERGMs represent the probability distribution of adjacency matrix $\bm{A}$ as
\begin{equation}
\label{eqn:ergm}
 \mathbb{P}_{\bm{\theta}_G}(\bm{A}= \bm{a}|\bm{X}_G) = \frac{\exp[\bm{g}(\bm{a}, \bm{X}_G)\cdot \bm{\theta}_G]}{\eta(\bm{\theta}_G, \mathcal{A})}, \quad \quad \bm{a} \in \mathcal{A},
 \end{equation}
where $\mathcal{A}$ is the support of $\bm{A}$, $\bm{g}(\bm{a}, \bm{X}_G)$ is a $q$-vector of sufficient statistics embodying local features of the social network which we allow to depend on exogenous covariates $\bm{X}_G$, $\bm{\theta}_G \in \bm{\Theta}_G \subset \mathbb{R}^q$ is a parameter vector, and the denominator $\eta(\bm{\theta}_G, \mathcal{A})$ is a normalizing constant defined as
$$\eta(\bm{\theta}_G, \mathcal{A}) = \sum_{a \in \mathcal{A}} \exp[\bm{g}(a, \bm{X}_G)\cdot \bm{\theta}_G]. $$
Model (\ref{eqn:ergm}) treats the random adjacency matrix $\bm{A}$ as the response in a regression model in which predictors are network statistics that are chosen to capture the main structural features of the network  \citep{hunter2008ergm, kolaczyk2014statistical}. For instance, the number of edges can be represented through the sufficient statistic $g_0(\bm{a}, \bm{X}_G) \equiv g_0(\bm{a}) = \sum_{i < j} a_{ij}= |\mathbb{E}|$. Loosely speaking, this method models the probability that a pair of nodes in a network will have a tie between them given all the other ties which are present. It can be shown that the $r$-th component of $\bm{\theta}_G$, $\theta_{G,r}$, may be interpreted as the increase in the conditional log-odds of the network per unit increase in the corresponding component of $\bm{g}(\bm{a}, \bm{X}_G)$ resulting from switching the value of $A_{ij}$ from 0 to 1 and holding all the other dyads fixed \citep{robins2007introduction, hunter2008ergm}.

It is possible to control for exogenous network effects by allowing the propensity of an edge joining two vertices to depend on nodal attributes. For instance, given a factor variable $\bm{X}_{G,l}$, $l \in \{1, \ldots, q\}$, and a level $m$, the nodal factor (NF) statistic counts gives the number of times a vertex with that factor level appears in
an edge in the network \citep{morris2008specification, hunter2008goodness}. The NF statistic is given by $g_1(\bm{a}, \bm{X}_{G}) = \sum_{i < j } a_{ij} h_1(\bm{X}_{G,i}, \bm{X}_{G,j})$, where
$$ h_1(\bm{X}_{G,i}, \bm{X}_{G,j}) = \begin{cases} 2 & \text{if both $i$ and $j$ have a specific factor level $m$}\\ 1 & \text{if either $i$ or $j$ (but not both) has a specific factor level $m$}\\
0  & \text{if neither $i$ nor $j$ has the specified factor level $m$}\end{cases}.$$ 
We can also define interaction terms for nodal attribute based mixing. Uniform homophily (UH), defined by $g_2(\bm{a}, \bm{X}_G) = \sum_{i < j } a_{ij} h_2(\bm{X}_{G,i}, \bm{X}_{G,j})$, where $ h_2(\bm{X}_{G,i}, \bm{X}_{G,j}) = \mathds{1}(\bm{X}_{G,li} = \bm{X}_{G,lj})$ is the indicator for equivalence of the categorical attribute $\bm{X}_{G,l}$, quantifies the extent to which similar nodes tend to form ties.

We may also expect the propensity to form friendships to depend on the difference between
two individuals’ values of an ordinal categorical variable. The absolute difference (AD) statistic is defined by $g_3(\bm{a}, \bm{X}_G) = \sum_{i < j } a_{ij} h_3(\bm{X}_{G,i}, \bm{X}_{G,j}),$ where $h_3(\bm{X}_{G,i}, \bm{X}_{G,j}) = |\bm{X}_{G,li} -  \bm{X}_{G,lj}|.$

 All of the above statistics correspond to \textit{dyadic independence terms}, as the associated variation in $\bm{g}(\bm{a}, \bm{X}_G)$ from switching the value of $A_{ij}$ from 0 to 1 may always be calculated without knowing anything about $\bm{a}$ (see Definition 2 in \citep{hunter2008ergm}). Models that contain only dyadic independence terms are termed dyadic independence ERGMs. More realistic models will capture high-order dependency structure in the network through dyadic dependence terms \citep{hunter2008ergm, morris2008specification}. We consider models that incorcoporate linear combinations of the entire distribution of degree or shared partner
statistics \citep{hunter2008ergm}. One such term is the geometrically weighted degree (GWD) statistic, which is defined as 
 $$ g_4(\bm{a}; \gamma) = \sum_{d=0}^{N - 1} e^{-\gamma d} N_d(\bm{a}),  $$
 where $N_d(\bm{a})$ is the number of nodes with degree $d$ in network $\bm{a}$. The parameter $\gamma$ is typically fixed by the user and controls the extent to which higher-degree nodes are likely to occur in $G$ \citep{kolaczyk2014statistical}. Proposed by Snijders et al. (2006), the geometrically weighted edgewise shared partner (GWESP) statistic can be expressed as
$$ g_5(\bm{a};\lambda) = e^{\lambda} \sum_{k=1}^{n-2} \left\{1- \left(1- e^{\lambda} \right)^i \right\} \text{EP}_k(\bm{a}),$$
where $\text{EP}_k(\bm{a})$ is the number of edges in adjacency matrix $a$ between two nodes that share $k$ neighbors \citep{snijders2006new, hunter2008ergm}. In other words, $\text{EP}_k(\bm{a})$ is the number of $k$-triangles in the network, where a $k$-triangle is the number of individual triangles sharing a common base, a triangle is a closed triplet, and a triplet is three nodes that are connected by either two (open triplet) or three (closed triplet) undirected ties \citep{kolaczyk2014statistical}. Thus, including this term in an ERGM can capture the level of transitivity or clustering in the network, which is the tendency for nodes connected to a certain node to also have connections between themselves. Other examples of statistics that capture higher-order global network structure include the geometrically weighted dyadwise shared partner (GWDSP), alternating $k$-stars, and alternating $k$-triangles statistics, which are discussed in \cite{snijders2006new, robins2007recent,morris2008specification}. The geometrically weighted terms are effective at overcoming degeneracy problems encountered with other models that incorporate \textit{dyadic dependence} such as Markov network models \citep{frank1986markov, snijders2006new, hunter2008ergm, kolaczyk2014statistical}.

Maximum likelihood inference for $\bm{\theta}_G$ is usually performed using Monte Carlo Markov Chain (MCMC) maximum likelihood estimation, a method implemented in the \texttt{ergm} package of the suite \texttt{statnet} in \texttt{R} \citep{hunter2008ergm, yauck2021sampling}. Pseudolikelihood estimation \citep{besag1974spatial}, which assumes independence of dyads, may be safely used with dyadic independence models \citep{hunter2008ergm}. Otherwise, MCMC maximum likelihood must be used since the normalization factor $\eta(\bm{\theta}_G, \mathcal{A})$ is mathematically intractable with the presence of higher-order dependence terms in the ERGM. 

\section{Methods}
\label{section:methods}
Our simulation approach consists of resampling from the observed covariate and exposure data to preserve the empirical associations among these variables. In the case of private nodal covariates, we suggest procedures for generating samples from a multivariate distribution that matches the empirical joint distribution in the original data.  
Outcomes are generated based on direct and indirect treatment effects and associations with pretreatment covariates estimated in the original data. An ERGM is fitted on the original data and then used to generate repeated random networks conditional on the simulated nodal attributes so as to retain essential characteristics of the original social network. Our approach conforms with a randomized response framework \citep{karwa2017sharing}, where a synthetic network $G^*$ is obtained by randomizing the presence/absence of a tie to each dyad based on the estimated ERGM parameters. Where appropriate, we also discuss special considerations that might arise with multilevel network studies, in which case the network may be decomposed in disjoint subgraphs (e.g., social networks of students embedded within schools). Figures \ref{fig:workflow} and \ref{fig:workflow2} summarize the synthetic plasmode simulation framework for the cases of public and private nodal attributes, respectively. 

\subsection{Construct the study population}
The first task in creating simulated networks is to create the study sample on which the simulations will be based from the larger dataset. Constructing the study sample for a network simulation study first requires to define the target population network. Particularities of the study design, including inclusion and exclusion criteria for the population, definitions of exposures and covariates may have an effect on the subsequent performance evaluation of statistical methods. For instance, computational aspects of the intended analyses to be performed on the dataset might lead an investigator to exclude isolates, i.e., nodes with degree 0. In the case of multilevel network studies, a subset of clusters may be considered for inclusion in the target population.

\subsection{Select covariates for simulation basis}


We recommend specifying a set of covariates $\bm{X}$ that are believed to be associated with the outcome and tie formation mechanisms, including important demographic information such as age, gender, and race. Recall that we refer to the subsets of variables used to generate the outcome variable and the graph as $\bm{X}_Y \subseteq \bm{X}$ and $\bm{X}_G \subseteq \bm{X}$, respectively. Including more covariates in $\bm{X}_Y$ and $\bm{X}_G$ may result in more realistic simulated outcomes and graphs. However, including all potential covariates in $\bm{X}_Y$ and $\bm{X}_G$ might be infeasible due to the model estimation procedures required in subsequent steps. If any of the variables in $\bm{X}_Y$ are associated with the exposure, then confounding will be present in the simulated datasets. Additionally, if there are variables in $\bm{X}_Y \cap \bm{X}_G$ that are also associated with exposure, there will be latent homophily confounding if these variables are not controlled for at the analysis stage.


\subsection{Estimate associations among the exposure and covariates}
\label{section:associationexposurecovariates}
The investigator may choose to skip this step if the nodal attributes are deemed public. More often than not, just like the tie variables, the nodal attributes will be private and will have to be synthetically created. We distinguish between two main classes of synthetic nodal data release mechanisms that provide protection of privacy: \begin{enumerate}
\item \textit{Direct release of synthetic data}: Given private nodal attributes, direct release of synthetic data consists of constructing a statistical model that captures the statistical properties of the real data and then to use this model to generate different synthetic nodal attributes that are statistically similar via stochastic simulation. A variety of methods can be used to this end \citep{dankar2021fake,el2020practical}. The Synthetic Data Vault (SDV) is a synthetic data generator implemented in Python which estimates the joint distribution of the population using a latent Gaussian copula \citep{patki2016synthetic}. Another generator implemented in R is the package \texttt{synthpop}, which generates a synthetic dataset sequentially one attribute at a time by estimating conditional distributions and can accommodate categorical and continuous variables \citep{synthpop}. As these issues are not the focus of this paper, we refer the reader to \citep{dankar2021fake,el2020practical} for a review of synthetic data generation techniques. 
\item \textit{Synthetic data generation based on the release of sufficient statistics}: For this release mechanism, we fit a statistical model on the input nodal dataset and release a set of sufficient statistics, denoted $\hat{\bm{\theta}}_{ZX}$, which capture the marginal distributions and the dependence structure observed among the private nodal attributes. Using the same statistical model, we can proceed to stochastic production of synthetic data. This approach might appear less restrictive if files need to be transferred by the data owner to a user from a virtual data enclave, as it entails transferring a vector of parameter estimates $\hat{\bm{\theta}}_{ZX}$ instead of whole synthetic datasets. This can be accomplished through copula-based synthesis techniques, such as methods implemented in the R package \texttt{VineCopula} \citep{vinecopula}. For categorical data, the R package \texttt{GenOrd} implements a Gaussian-copula based method which allows the user to simulate synthetic data whose structure matches that observed in the real dataset only from two inputs: the empirical marginal distributions and correlation matrix of the categorical private nodal covariates. One important caveat is that copula-based synthesis techniques might be infeasible when faced with mixtures of continuous and discrete variables, which are prevalent in real-world datasets.  
\end{enumerate}

\subsection{Specify and estimate a model for the outcome}
In order to produce simulated outcomes that have realistic associations with the exposure and pretreatment covariates, we specify and estimate an outcome model in the real dataset. For instance, in the case of a continuous outcome, Model (\ref{eqn:modely}) might be appropriate if interference effects are of interest to the investigator. Recall that this model is conditional on the sociomatrix $\bm{A}$ through the neighborhood exposure effect. Researchers can tailor this model to capture important features of how covariates and exposure relate to the outcome, such as nonlinearity and treatment heterogeneity. For instance, Model (\ref{eqn:modely}) could further be enriched with interaction terms between pretreatment covariates and exposure to incorporate effect heterogeneity. In the case of multilevel network data, cluster-level random effects could be incorporated to capture within-cluster dependence in the outcome data. We denote the estimated parameter vector for the outcome model as $\hat{\bm{\theta}}_Y$, which might include, for example, estimated coefficients $\hat{\bm{\beta}}$ and the estimated residual variance $\hat{{\sigma}}_{\varepsilon}^2$.  

\subsection{Specify and estimate an exponential random graph model}

In order to release synthetic networks that resemble the graph under consideration, we must specify an ERGM that fits the data well. In an ERGM, the predictor variables are direct functions of the response variable (i.e., the state of a tie between each pair of nodes) such that these models can be thought of as autoregressive or autologistic models \citep{besag1974spatial, morris2008specification}. This has an impact on many aspects of specification and estimation of ERGMs. We cannot adequately cover the subject of building ERGMs given the limited space available and limited scope of this article, so we refer the reader to seminal papers in the ERGM literature.  Morris et al.~(2008) describe the classes of statistics that are available in the \texttt{ergm} package, from basic terms to nodal attribute effects and curved exponential family terms \citep{morris2008specification}.

Past experimentation with ERGMs has shown that random graphs simulated from ERGMs with the maximum likelihood estimator can bear little resemblance with the original network \citep{handcock2003assessing, hunter2008goodness}. This troubling fact might be explained by the fact that although the MLE places a relatively high probability on the observed network, this probability might be extremely small relative to other networks \citep{hunter2008goodness}. Problems of degeneracy encountered with ERGMs \citep{frank1986markov} can be alleviated through the use of GWD and GWESP terms introduced in Section \ref{section:prelim_ERGM} \citep{snijders2006new, hunter2008goodness}. Exogenous nodal attribute effects can also play an important role in the generative processes that give rise to networks \citep{hunter2008goodness}. Model building should be performed in conjunction with goodness-of-fit assessments via Monte Carlo Markov Chain simulation to ensure that ERGMs provide good fit to the data, as described by Hunter et al.~(2008) \citep{hunter2008goodness}.

Multilevel network studies may require special handling. We first recommend fitting an ERGM to each disjoint subgraph separately, as is usually done in social network analysis when faced with disconnected subgraphs \citep{hunter2008goodness, kolaczyk2014statistical}. The question then arises as to whether a single ERGM should be specified for all the clusters or if the specification should be tailored to each cluster separately. If the same, single specification is used for all disjoint subgraphs, then ideally the nodal covariate distribution is sufficiently similar in all clusters. Otherwise, the ERGM parameters may vary from one subgraph to the other when the support of exogenous nodal attributes included in the model differs across subgraphs. For instance, a given level of a categorical nodal attribute might be represented in one subgraph and not be present at all in another. This can even lead to the model being overspecified (i.e., containing redundant terms) in some subgraphs, if, for example, the model includes a term for differential homophily \citep{hunter2008goodness}. Alternatively, instead of assuming that all subgraphs originate from the same generative model, one could specify and fit a different ERGM on each disjoint subgraph, in which case the aforementioned problems would be avoided. Alternatively, instead of assuming that all subgraphs originate from the same generative model, one could specify and fit a different ERGM on each disjoint subgraph, in which case the aforementioned problems would be avoided.

\subsection{Resample and simulate}

We construct $S$ simulated datasets of size $n \leq N$, where $N$ is the size of the full sample. We describe the procedure for simulating one dataset, and this process is repeated $S$ times. If the nodal attributes are public (i.e., not subject to privacy concerns), we sample $n$ units in the study population with replacement to form a bootstrap sample of nodal attributes $(Z^*, \bm{X}^*)$. If the nodal attributes are private, we can generate a sample of $n$ vectors of synthetic nodal attributes $(Z^*, \bm{X}^*)$ using the method described in Section \ref{section:associationexposurecovariates}. Using the simulated nodal attributes in $\bm{X}^*_G \subseteq \bm{X}$ and the estimated ERGM parameters $\hat{\bm{\theta}}_G$, a random graph $G^*$ may be simulated via a Monte Carlo Markov Chain method implemented in the \texttt{ergm} package. We denote the associated simulated sociomatrix as $\bm{A}^*$. Finally, we may generate outcomes $Y^*$ stochastically from the same model that was fitted on the original data, which may be conditional on the simulated graph structure $G^*$, the simulated exposure $Z^*$, and the pretreatment covariates $\bm{X}^*_Y \subseteq \bm{X}^*$. 

\subsection{Analyze simulated data}

We now have $S$ simulated network datasets of size $n$, each of which contains values of the exposure $Z^*$, values of the pretreatment treatment covariates $\bm{X}^*$, values of the outcome $Y^*$ for all nodes in the network, and the associated graph structure $G^*$. The complete generating mechanism is known for all the aforementioned data structures. In the case of public nodal attributes, the data generating mechanism for $Z^*$ and $\bm{X}^*$ is unknown because these data are unaltered from their observed values, preserving any associations that exist among them in the observed data.

If desired, unmeasured confounding can be induced at the analysis stage by setting aside predictors of the outcome in $\bm{X}^*_Y$ from the analysis to be unobserved confounders in $\bm{U}^*$. We may observe the performance of methods under unobserved confounding by hiding the variables in $\bm{U}^*$ from the confounding adjustment methods applied to the simulated data. If some variables in $\bm{U}^*$ are also in $\bm{X}^*_G$, then latent homophily confounding could ensue since the withheld variables are also linked to the tie formation mechanism.

Analyzing these data, we return estimates of relevant causal contrasts for each of the $S$ simulated
datasets. Using these estimates, we may evaluate finite-sample properties (e.g., bias, variance) of the estimation
procedure as is usually done in typical simulation studies.

\begin{figure}
\tikzset{arrow/.style={-stealth, thick, draw=gray!80!black}}
\begin{tikzpicture}[ampersand replacement=\&]

    \node[text width=5cm] at (-1.5, 5) (A) {\textsc{Create sample \& \\ select covariates}};

    \node[text width=3cm] at (2, 5) (C) {\textsc{Estimate\\ models}};

    \node[text width=3cm] at (4.3, 5) (D) {\textsc{Resample}};

    \node[text width=3cm] at (7, 5) (D) {\textsc{Simulate graphs \&\\ outcomes}};
    
    \node[text width=5cm] at (10.5, 5) (G) {\textsc{Analyze\\data}};

    \node at (-4.1, 4.2) (H) {};
    \node at (10, 4.2) (I) {};
    \path (H) edge (I);

    \node at (-3, 3.5) (J) {$Y$};
    \node at (-3, 2.5) (K) {$Z$};
    \node at (-3, 1) (L) {$\bm{X}$};
    \node at (-3, -0.5) (M) {$G$};

    \node at (-1, 1.3) (P) {$\bm{X}_Y$};
     \node at (-1, 0.7) (Q) {$\bm{X}_G$};

    \path[dashed] (L) edge (P);
    \path[dashed] (L) edge (Q);

    \node[rectangle, rounded corners, draw, fill=blue!20, minimum height=1cm] at (1.2,3.5) (S) {$\hat{\bm{\theta}}_Y$};

\path[arrow, bend left] (M) edge (S);
    \path[arrow] (J) edge (S);
    \path[arrow] (K) edge (S);
    \path[arrow] (P) edge (S);

    \node[rectangle, rounded corners, draw, fill=cyan!20, minimum height=1cm] at (1.2,-0.5) (T) {$\hat{\bm{\theta}}_G$};


    \path[arrow] (Q) edge (T);
    \path[arrow] (M) edge (T);


    \node at (3.8, 2.5) (X) {$(Z_1^*, \ldots, Z_S^*)$};
    \node at (3.8, 1.3) (Y) {$(\bm{X}^*_{1Y}, \ldots, \bm{X}_{SY}^* )$};
    \node at (3.8, 0.7) (Z) {$(\bm{X}^*_{1G}, \ldots, \bm{X}_{SG}^* )$};

    \path[dashed] (P) edge (Y);
    \path[dashed] (Q) edge (Z);

    \node at (6.6, -0.5) (A1) {$(G^*_1, \ldots, G^*_S)$};
    
     \node at (6.6, 3.5) (A2) {$(Y^*_1, \ldots, Y^*_S)$};
\path[arrow,bend left]  (A1) edge (A2);
    \path[dashed] (K) edge (X);

    \node at (3.8, 2.5) (A3) {$(Z_1^*, \ldots, Z_S^*)$};

    \path[arrow] (A3) edge (A2);

    \path[arrow] (S) edge (A2);
    \path[arrow] (T) edge (A1);

     \node[rectangle, rounded corners, draw, fill=red!20, minimum height=1cm] at (9,1.5) (A4) {$(\hat{\bm{\theta}}_1^*, \ldots, \hat{\bm{\theta}}^*_S)$};

    \path[arrow] (A2) edge (A4);
    \path[arrow] (A1) edge (A4);
    \path[arrow] (A3) edge (A4);
    \path[arrow] (Y) edge (A2);
    \path[arrow] (Z) edge (A1);
    \path[arrow] (Y) edge (A4);

    \node at (-4.1, -1.2) (A7) {};
    \node at (10, -1.2) (A8) {};
    \path (A7) edge (A8);
    
    
    
    
     
\end{tikzpicture}
\caption{Diagram showing the steps in the simulation framework for the case of \textit{public nodal attributes}. Dashed lines represent reusing or resampling a data element without modification. Solid arrows represent creating new data structures from an existing data element.}
\label{fig:workflow}
\end{figure}
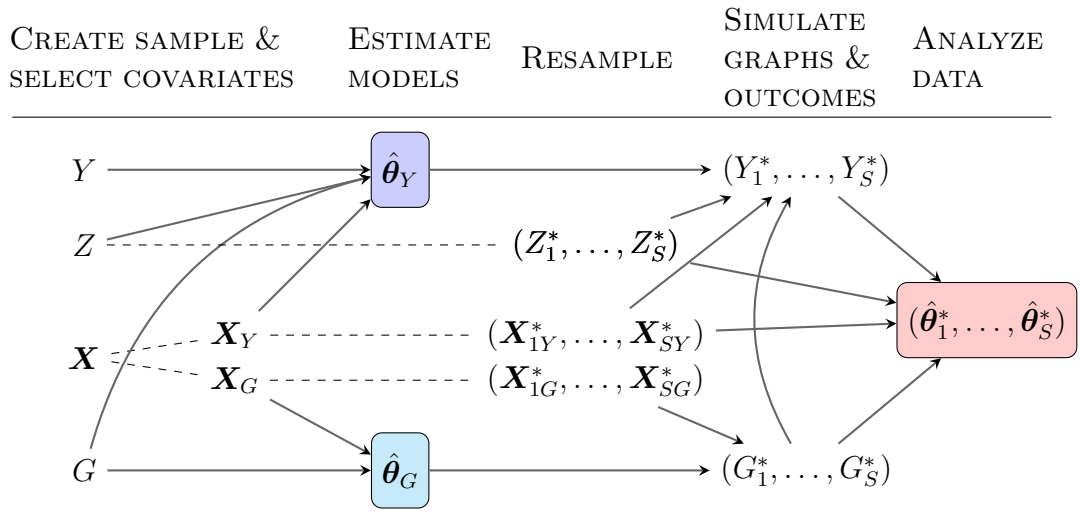

\begin{figure}
\tikzset{arrow/.style={-stealth, thick, draw=gray!80!black}}
\begin{tikzpicture}[ampersand replacement=\&]

    \node[text width=5cm] at (-1.5, 5) (A) {\textsc{Create sample \& \\ select covariates}};

    \node[text width=3cm] at (2, 5) (C) {\textsc{Estimate\\ models}};

    \node[text width=3cm] at (4.3, 5) (D) {\textsc{Simulate \\exposure \&\\ covariates}};

    \node[text width=3cm] at (7, 5) (D) {\textsc{Simulate graphs \&\\ outcomes}};
    
    \node[text width=5cm] at (10.5, 5) (G) {\textsc{Analyze\\data}};

    \node at (-4.1, 4.2) (H) {};
    \node at (10, 4.2) (I) {};
    \path (H) edge (I);

    \node at (-3, 3.5) (J) {$Y$};
    \node at (-3, 2.5) (K) {$Z$};
    \node at (-3, 1) (L) {$\bm{X}$};
    \node at (-3, -0.5) (M) {$G$};

    \node at (-1, 1.3) (P) {$\bm{X}_Y$};
     \node at (-1, 0.7) (Q) {$\bm{X}_G$};

    \path[dashed] (L) edge (P);
    \path[dashed] (L) edge (Q);

    \node[rectangle, rounded corners, draw, fill=blue!20, minimum height=1cm] at (1.2,3.5) (S) {$\hat{\bm{\theta}}_Y$};

\path[arrow, bend left] (M) edge (S);

    \path[arrow] (J) edge (S);
     \path[arrow] (K) edge (S);
    \path[arrow] (P) edge (S);

    \node[rectangle, rounded corners, draw, fill=cyan!20, minimum height=1cm] at (1.2,-0.5) (T) {$\hat{\bm{\theta}}_G$};

    \node[rectangle, rounded corners, draw, fill=green!20, minimum height=1cm] at (1.2,1.9) (Z1) {$\hat{\bm{\theta}}_{ZX}$};
      \path[arrow] (K) edge (Z1);
      \path[arrow] (P) edge (Z1);
      \path[arrow] (Q) edge (Z1);


    \path[arrow] (Q) edge (T);
    \path[arrow] (M) edge (T);


    \node at (3.8, 2.5) (X) {$(Z_1^*, \ldots, Z_S^*)$};
    \node at (3.8, 1.3) (Y) {$(\bm{X}^*_{1Y}, \ldots, \bm{X}_{SY}^* )$};
    \node at (3.8, 0.7) (Z) {$(\bm{X}^*_{1G}, \ldots, \bm{X}_{SG}^* )$};

    \path[arrow, bend left] (Z1) edge (X);
      \path[arrow] (Z1) edge (Y);
      \path[arrow, bend right] (Z1) edge (Z);


    \node at (6.6, -0.5) (A1) {$(G^*_1, \ldots, G^*_S)$};
     \node at (6.6, 3.5) (A2) {$(Y^*_1, \ldots, Y^*_S)$};
\path[arrow,bend left] (A1) edge (A2);

    \node at (3.8, 2.5) (A3) {$(Z_1^*, \ldots, Z_S^*)$};

    \path[arrow] (A3) edge (A2);

    \path[arrow] (S) edge (A2);
    \path[arrow] (T) edge (A1);

     \node[rectangle, rounded corners, draw, fill=red!20, minimum height=1cm] at (9,1.5) (A4) {$(\hat{\bm{\theta}}_1^*, \ldots, \hat{\bm{\theta}}^*_S)$};

    \path[arrow] (A2) edge (A4);
    \path[arrow] (A1) edge (A4);
    \path[arrow] (A3) edge (A4);
    \path[arrow] (Y) edge (A2);
    \path[arrow] (Z) edge (A1);
    \path[arrow] (Y) edge (A4);

    \node at (-4.1, -1.2) (A7) {};
    \node at (10, -1.2) (A8) {};
    \path (A7) edge (A8);
    
    
    
    
     
\end{tikzpicture}
\caption{Diagram showing the steps in the simulation framework for the case of \textit{private nodal attributes}. Note the key differences from Figure \ref{fig:workflow} are the steps which estimate $\bm{\theta}_{ZX}$ and generate synthetic covariate and treatment data from these estimated parameters. Dashed lines represent reusing or resampling a data element without modification. Solid arrows represent creating new data structures from an existing data element.}
\label{fig:workflow2}
\end{figure}
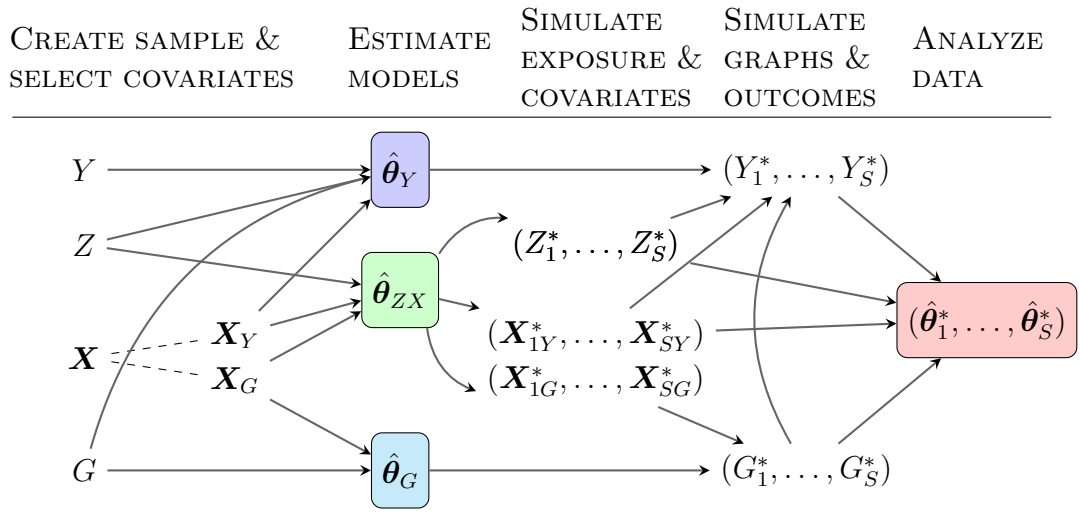

\section{Application}
\label{section:application}
\subsection{Study population}
We applied our framework to a cross-sectional network-based observational study of the spillover effects of maternal education on adolescent academic school performance. Our data come from the first wave of the National Longitudinal Study of Adolescent Health (Add
Health), a study of a nationally representative sample of 90,118 adolescent students in grades 7-12 in the
United States in 1994-95 who were followed through adolescence and the transition to adulthood \citep{harris2013add}. All participants were invited to take the In-School Survey during 1994 and 1995. Information on socio-demographic background, grades, school attendance, education and occupation of parents, extracurricular activities (e.g., club participation), and health status were collected. In addition to these, each student was asked to nominate up to five best female friends and five best male friends.

We examine an undirected network based on the data on friendship nominations from the first wave of data collection. We define a friendship as a symmetric relationship, meaning that there is an edge between students $i$ and $j$ if student $i$ listed $j$ as a friend in the in-school survey, or student $j$ listed $i$ as a friend, or both. This feature of reciprocation of nomination is common to many analyses of the Add Health network \citep{hunter2008goodness}. The students who did not report living with their biological mother, stepmother, foster mother, or adoptive mother at the time of the survey were excluded from the analytic sample because of the specific focus of the motivating example on the impact of maternal education. For computational considerations, we also excluded isolates from the analytic sample. This initial preparation resulted in an undirected network of 139 schools and 73,580 nodes in total. 

\begin{figure}[H]
\centering
	\begin{minipage}{0.45\linewidth}
	     \includegraphics[width=0.8\linewidth]{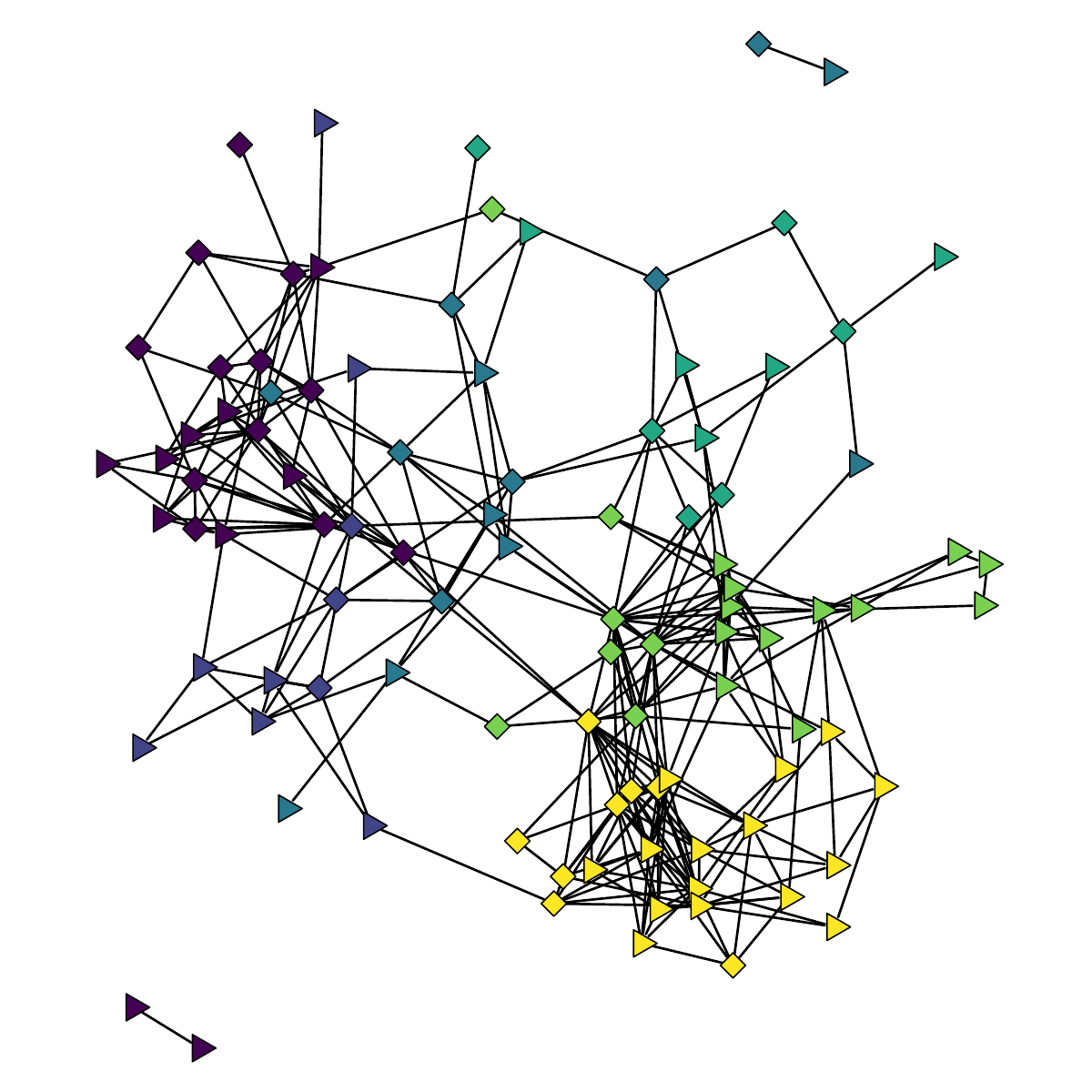}
	\end{minipage}
	\begin{minipage}{0.45\linewidth}
	    \includegraphics[width=0.8\linewidth]{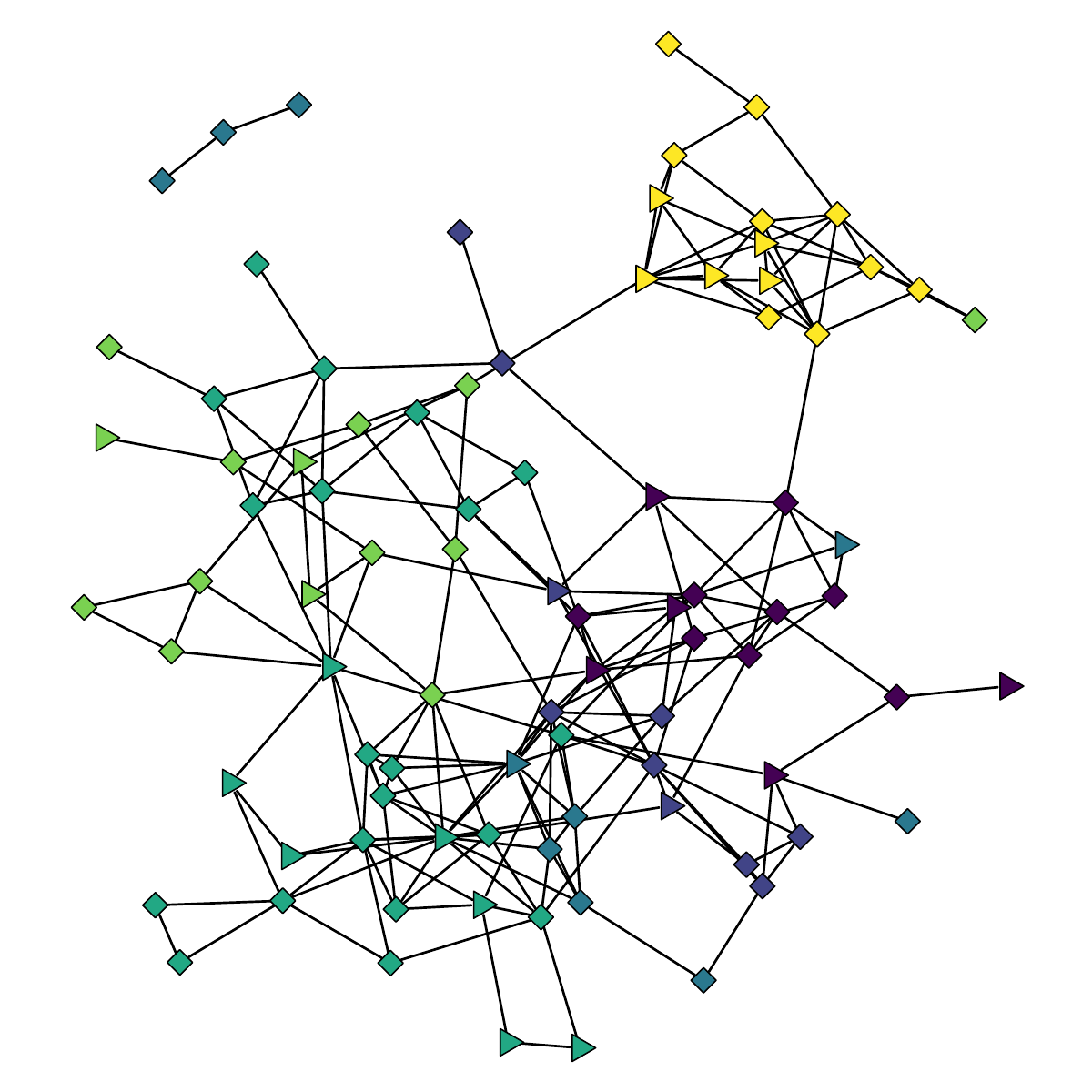}
        \end{minipage} \\	
	\begin{minipage}{0.45\linewidth}
	    \includegraphics[width=0.8\linewidth]{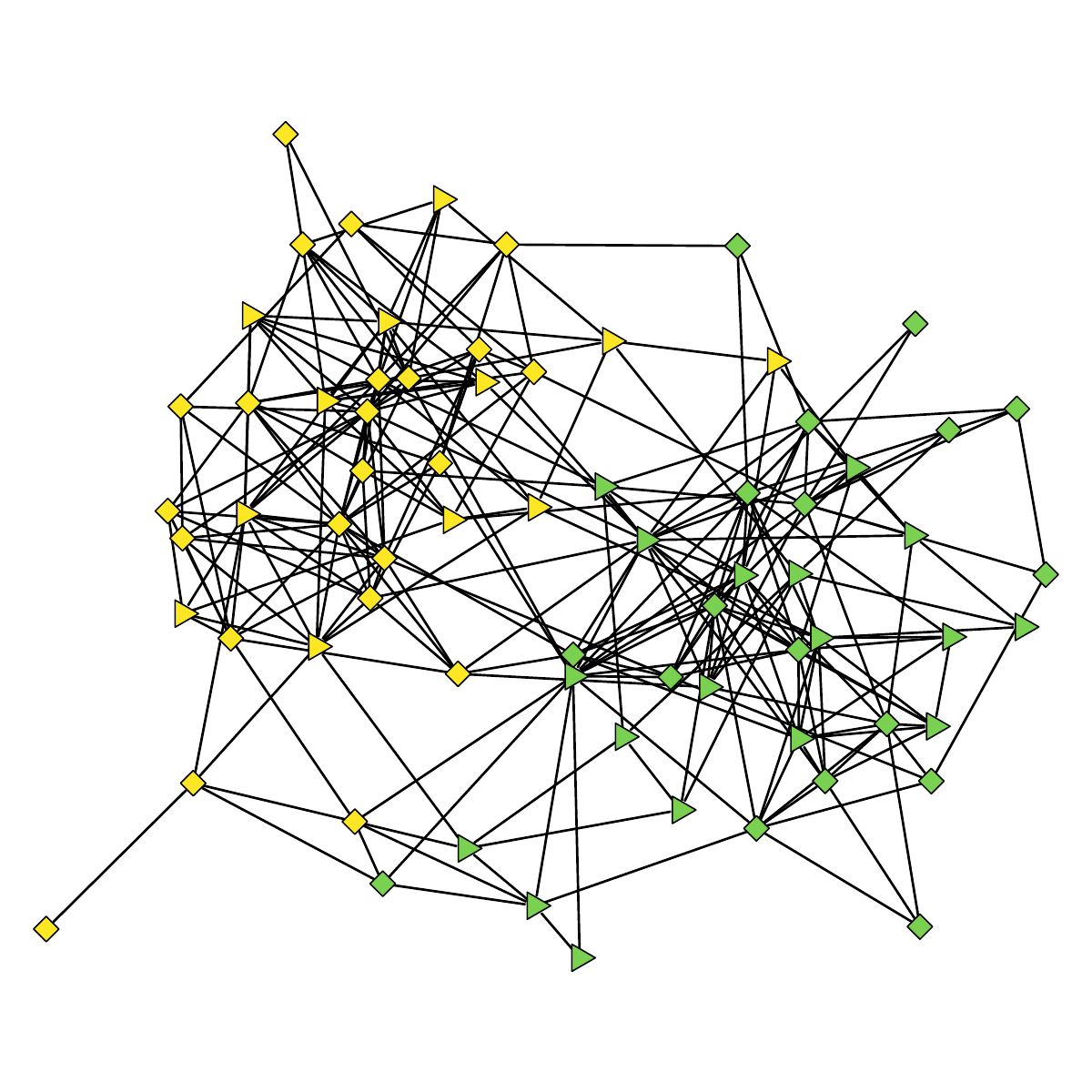}
         \end{minipage}
         \begin{minipage}{0.45\linewidth}
	    \includegraphics[width=0.8\linewidth]{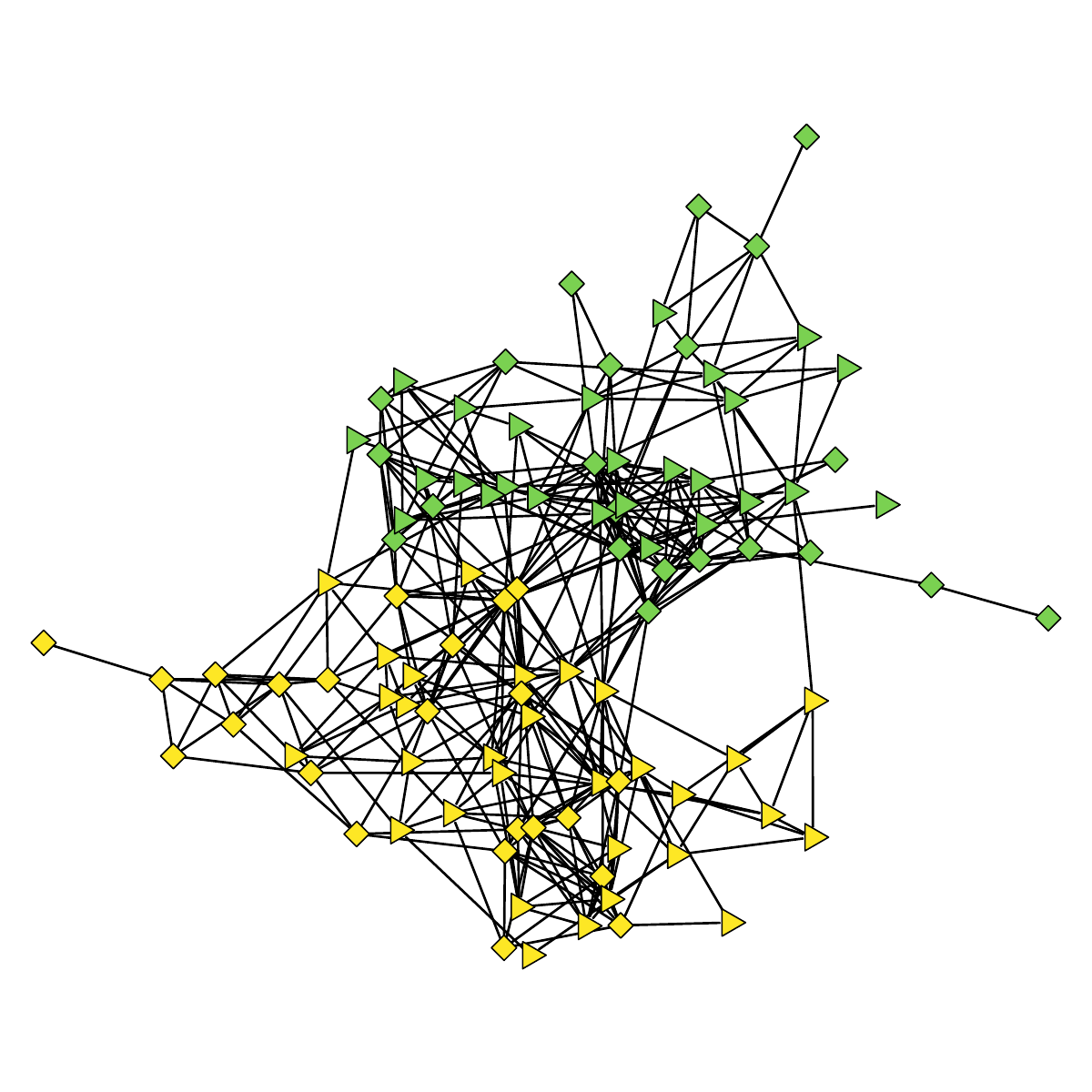}
         \end{minipage}\\	
	\begin{minipage}{0.45\linewidth}
	    \includegraphics[width=0.8\linewidth]{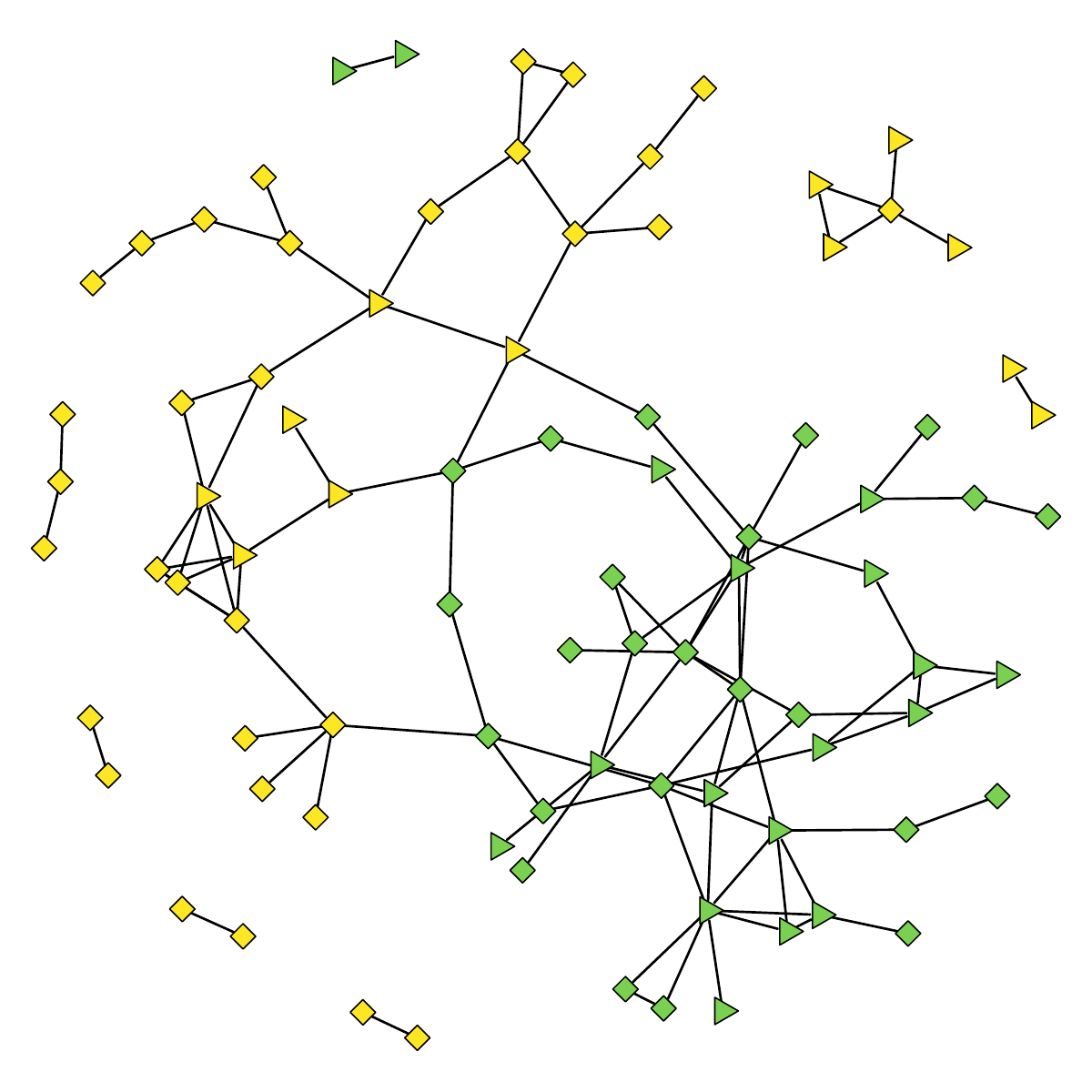}
         \end{minipage}
         \begin{minipage}{0.45\linewidth}
	    \includegraphics[width=0.5\linewidth]{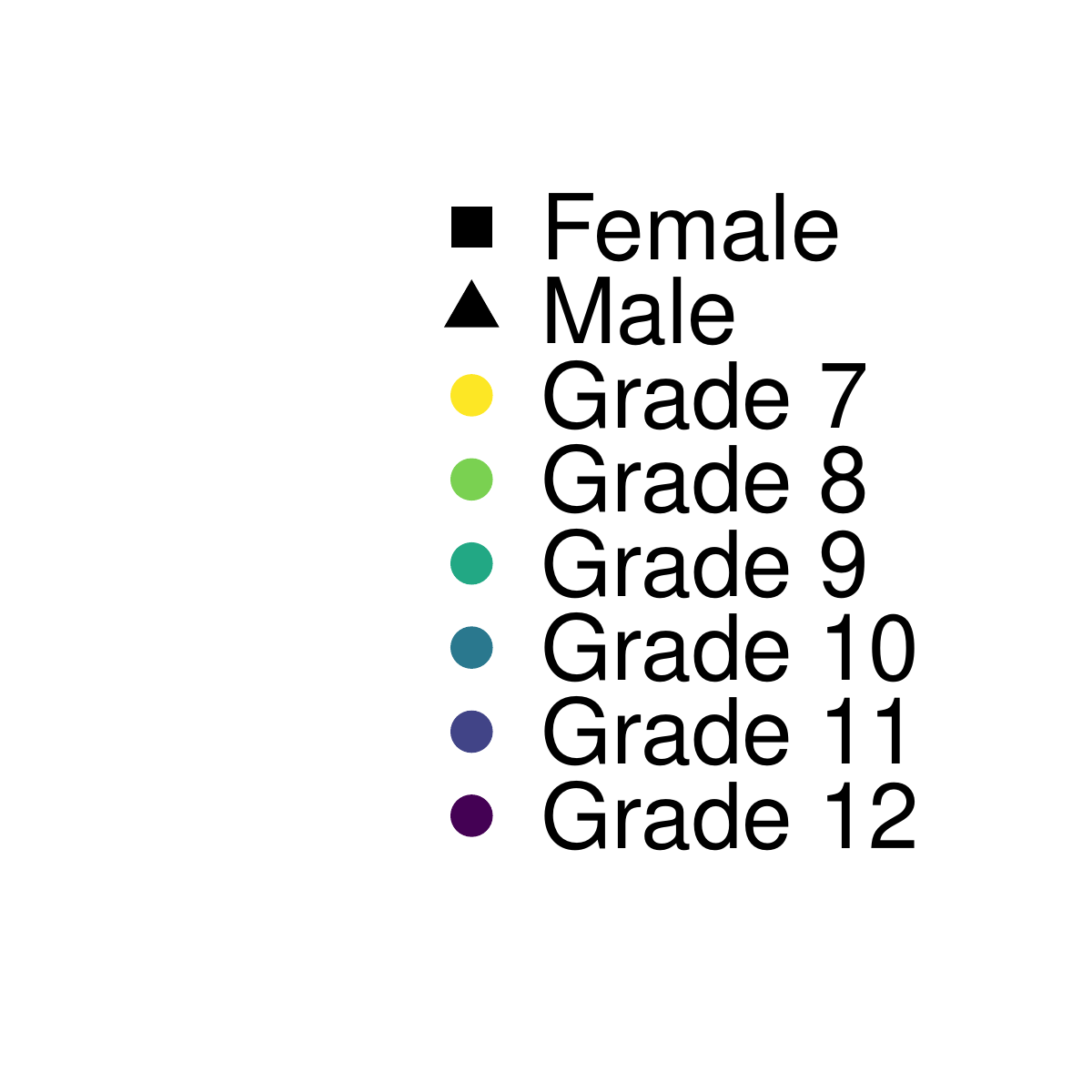}
         \end{minipage}
\caption{Mutual friendships the five selected schools in Add Health. From left to right, top to bottom, the networks that are shown correspond to School 003, School 028, School 106, School 122, and School 173. Shapes of nodes denote sex: squares for female and circles for males. The color indicates the grade (7 through 12).}
\label{fig:schools}
\end{figure}

To illustrate the simulation framework, we restricted ourselves to five schools whose size did not exceed 101 vertices. Figure \ref{fig:schools} displays the five networks under consideration in this study: School 003, School 028, School 106, School 122, and School 173. These five schools will form the basis of the simulation study and will be replicated as needed to increase the sample size. In what follows, the school type is indexed by $k$, $k=1, \ldots, 5$. \ref{appendix:networkcharacteristics} provides a description of the network characteristics for each of the five schools included in the study.




\subsection{Model specifications}

Since the nodal attributes in Add Health are subject to privacy concerns, not only the outcome and network ties had to be simulated, but the exposure and covariates as well. In such case, in accordance with the workflow shown in Figure \ref{fig:workflow2}, the exposure and covariates are generated based on the dependence structure and marginal distributions observed in the original dataset. Given that the exposure and covariates in the original dataset were categorical, we used the package \texttt{GenOrd} which implements a Gaussian copula method to generate the exposure and the covariates from a multivariate discrete distribution given pre-specified marginal distributions and bivariate correlation matrix \citep{barbiero2017r, yauck2021sampling}. However, this approach restricted the number of covariates that could be considered in the plasmode simulation, since we had to choose a set of covariates that yielded a feasible empirical correlation matrix given the empirical marginal distributions \citep{barbiero2017r}. In addition to the exposure, we considered the following set of categorical covariates: sex, race, school grade, whether the father is at home, screen time, motivation at school, sense of belonging, and physical fitness. A description of the variables retained for the simulation study can be found in \ref{appendix:variables}.



In order to create synthetic networks that resemble the five schools under consideration, we must specify an ERGM that provides a reasonable fit to the data. To this end, we explored different model specifications discussed in \citep{hunter2006inference} that were tested on different schools in Add Health. For all schools, we considered a dyadic dependence model which included 1) an edge count term; 2) a GWESP term with $\lambda =1 $; 3) a GWD term with $\lambda =1 $; 4) node factor (NF) terms for sex, grade, race; and 5) uniform homophily (UH) terms for sex, race, and grade. For Schools 003 and 028, in which there were six grades, we additionally included an absolute difference (AD) term for grade. The estimated ERGM coefficients for each school considered, along with visual assessments of the goodness of fit, can be found in Appendices C to G. In general, we see that the ERGMs do a respectable job of capturing features of the original networks, as simulated networks from the models resemble the schools in terms of degree, dyad-wise shared partners, and edge-wise shared partners distributions.


\subsection{Simulation}

\begin{figure}[h]

\centering
	\begin{minipage}{0.45\linewidth}
	    \includegraphics[width=1\linewidth]{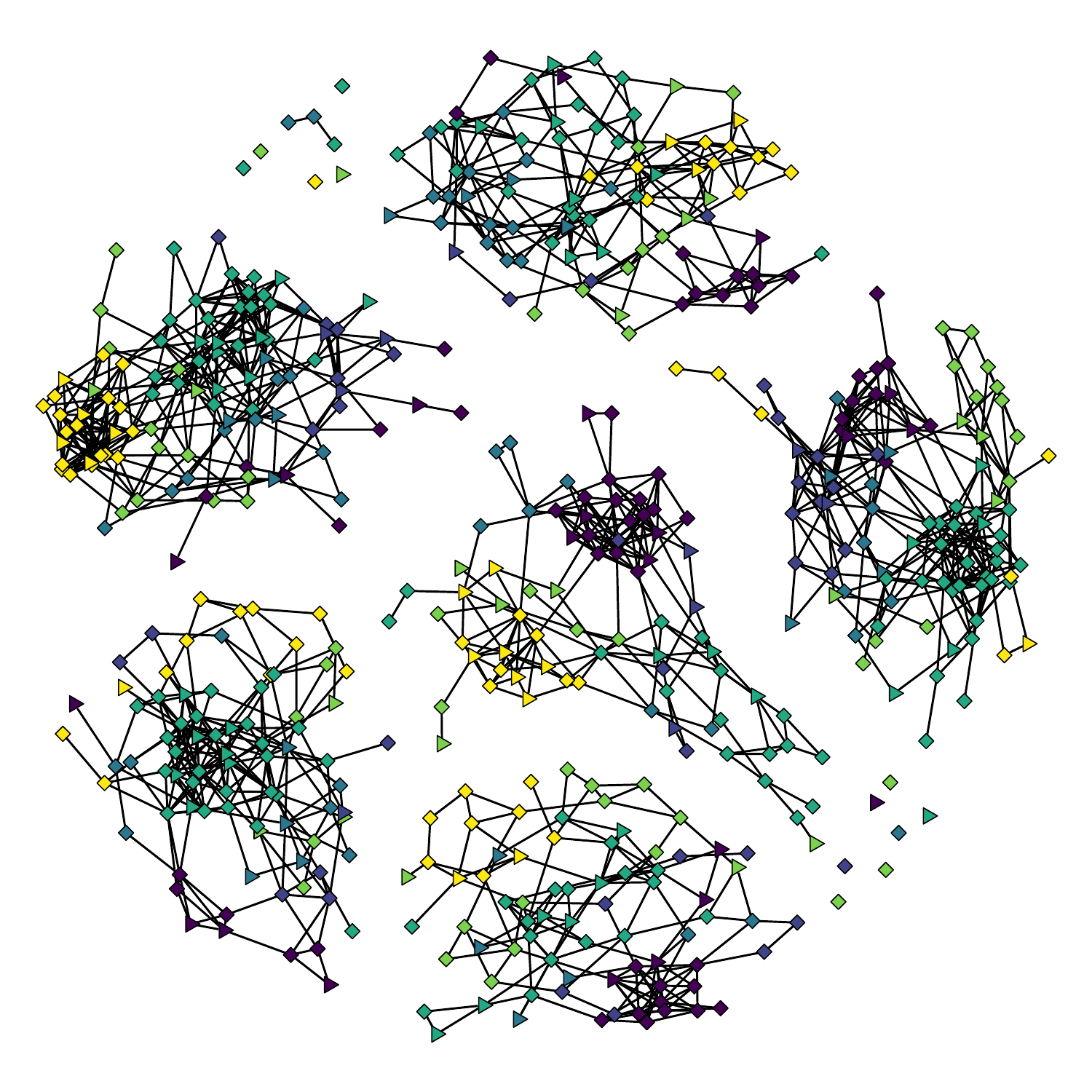}
	\end{minipage}
	\begin{minipage}{0.45\linewidth}
	    \includegraphics[width=1\linewidth]{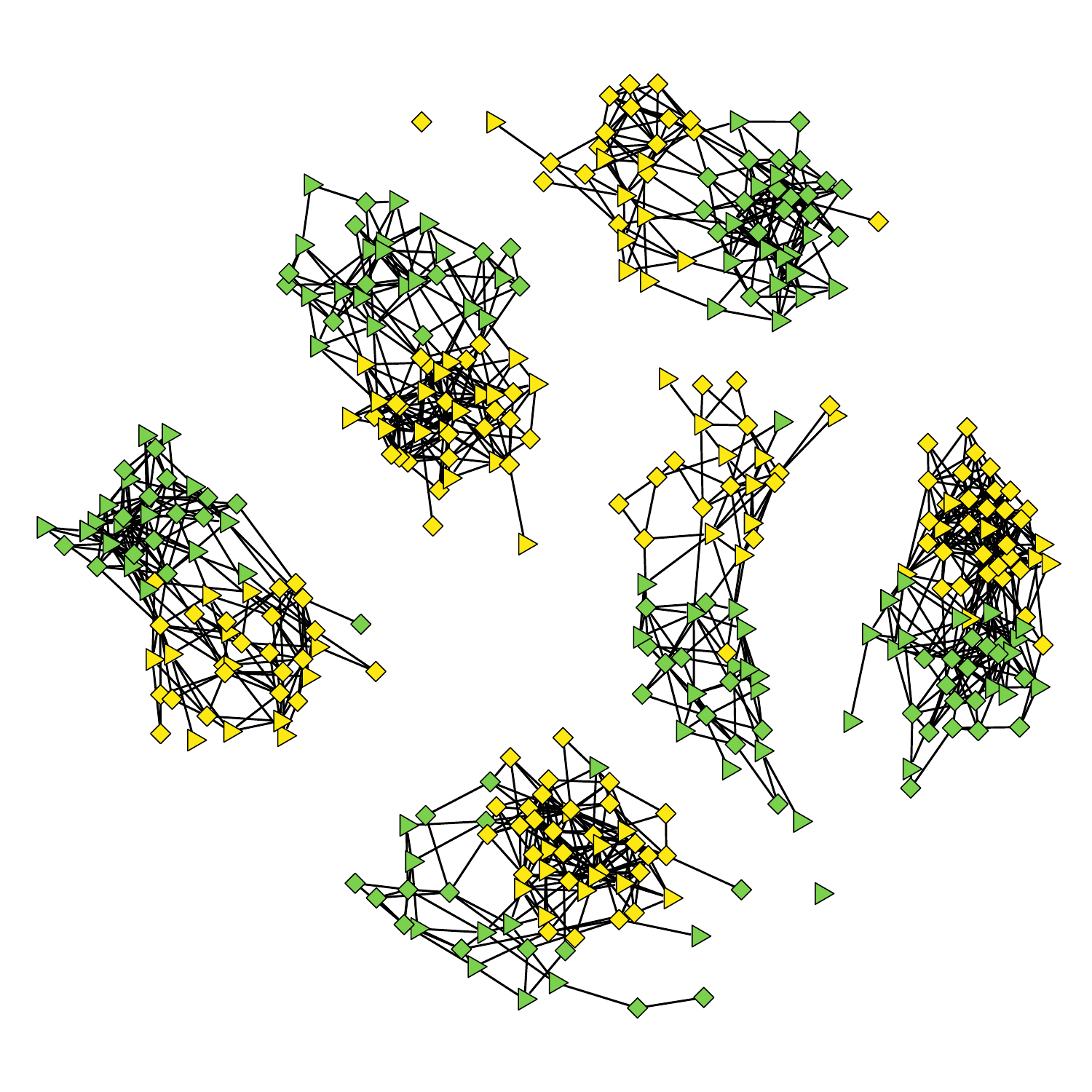}
        \end{minipage}\\	
\vspace*{-1.5cm}
\includegraphics[scale=0.18]{legend.pdf}
\caption{\textit{Left}: Six randomly generated networks from the fitted ERGM for School 028. \textit{Right}: Six randomly generated networks from the fitted ERGM for School 106.}
\label{fig:generatednetworks}
\end{figure}

For the purpose of evaluating previously proposed causal inference methods, we used the estimated model parameters and sufficient statistics derived in the previous step and expanded on the population of five schools by replicating units as needed. We considered a synthetic population of 30 schools where each school in the basis was replicated six times. The size of the $\nu$-th school of type $k$, denoted $N_{k \nu}$, was drawn from the distribution $\mathrm{Poisson}(N_{k})$, where $N_k$ is the size of the school in the original data (see Table \ref{tab:networkcharacteristics} for the original school sizes).  
We considered two schemes for the generation of the exposure:
\begin{enumerate}
\item In one scenario, the empirical bivariate associations between the exposure and the pretreatment covariates were preserved by simulating $Z^*$ and $\bm{X}^*$ using the package \texttt{GenOrd} with the empirical marginal distributions and Spearman rank correlation matrices provided for each school in Appendices C to G.
\item In a second scenario, only the empirical bivariate associations among the pretreatment covariates were preserved. Those were simulated through \texttt{GenOrd} while the exposure of node $i$, $i=1,\ldots, N_{k\nu}$, in the $\nu$-th school of type $k$ was generated according to the following model
\begin{equation}
\label{eqn:exposuregeneration}
\mathrm{logit}[\mathbb{P}(Z_{k\nu i}^* = 1)] = \gamma_0 + \bm{X}_{Z, {k \nu i}}^{*\top} \bm{\gamma} + b_{Z, k\nu}^*, 
\end{equation}
where $\bm{X}_Z^*$ denotes the set of simulated pretreatment covariates included in the exposure model and $b_{Z,k\nu}^* \sim N(0, \sigma^2_{b_Z})$ is a school-level random intercept. The exposure model included the student's race and the indicator of whether the adolescent's father is at home as pretreatment covariates. The parameter values were based on model estimates derived from fitting a mixed effects logistic regression model on the entire eligible sample ($N=73580$, 139 schools). The model terms and parameter values can be found in \ref{appendix:exposuremodel}.
\end{enumerate}
As mentioned above, the ERGM simulation process is conditional on the simulated exogenous nodal attributes $\bm{X}_G^*$. For a given school $\nu$ of type $k$, we computed network statistics corresponding to the NF, UH, and, where applicable, AD terms from the simulated pretreatment covariates in $\bm{X}_{G,k}^*$ and supplied the coefficients of the ERGM corresponding to the school of type $k$ to the function \texttt{ergm} to simulate the network ties. Figure \ref{fig:generatednetworks}
shows six randomly generated networks from the estimated ERGMs for Schools 028 and 106 conditional on the simulated exogenous attributes $\bm{X}^*_G$. It is worth mentioning that although we excluded isolates from the original network, we see a few isolates in the simulated network because of random variation. Although the presence of a few isolates does not impact our analyses, we could have prevented this by including a statistic controlling the number of isolates when specifying the ERGM.  

Lastly, we generated the standardized GPA for node $i$ in the $\nu$-th school of type $k$ as
\begin{multline}
\label{eqn:modelY2}
Y^*_{k \nu i} =  \beta_0 + \beta_Z Z_{k \nu i}^* + \beta_{\bm{Z}_{\mathcal{N}}} \frac{\sum_{j \in \mathcal{N}_{k \nu i}} Z^*_{k\nu j}}{d_{k \nu i}} \\ +\beta_{Z \times \bm{Z}_{\mathcal{N}}}Z^*_{ k \nu i} \frac{\sum_{j \in \mathcal{N}_{k \nu i}} Z^*_{k \nu j}}{d_{k \nu i}}+  \bm{X}_{Y, k\nu i} ^{*\top} \bm{\beta}_{\bm{X}_Y} + b_{Y, k\nu}^* + \varepsilon_{k \nu i},
\end{multline}
where $b_{Y,k\nu}^* \sim N(0, \sigma^2_{b_Y})$ is a school-level random intercept and $\varepsilon_{k\nu i} \sim N(0, \sigma^2_{\varepsilon})$ is measurement error. The parameter values for the outcome model can be found in \ref{appendix:outcomemodel}.

This simulation process was repeated 500 times, such that $S=500$ datasets each comprising 30 schools were considered for the evaluation of causal inference methods detailed in the next section. While network ties varied across simulation replicates, we maintained a fixed
total sample size of $N = 2,633$ nodes.

 
\subsection{Evaluation of estimators of interference effects in simulated data}
To illustrate the proposed simulation framework, we used the simulated data from both scenarios to compare estimators of the direct and indirect effects of the exposure. For each simulated dataset, using the definitions of the individual average potential outcome in (\ref{eqn:indapo}) and individual direct and indirect effects in (\ref{eqn:de}) and (\ref{eqn:ie}), respectively, we first computed the true population average causal effects defined as
$$DE(\alpha) = \frac{1}{30}\sum_{k=1}^5 \sum_{\nu=1}^6 \frac{1}{N_{k\nu}} \sum_{i=1}^{N_{k\nu}} DE_i(\alpha)$$  \text{and} $$ IE(\alpha, \alpha') = \frac{1}{30}\sum_{k=1}^5 \sum_{\nu=1}^6 \frac{1}{N_{k\nu}} \sum_{i=1}^{N_{k\nu}} IE_i(\alpha, \alpha'),$$
respectively. These population average causal effects are essentially averages of school-level averages \citep{mcnealis2023doubly}. For each node, the observed outcome was set to $Y_{k\nu i}=y_{k\nu i}(Z_{k\nu i}, \Sigma \bm{Z}_{\mathcal{N}_{k\nu i}})$. We applied inverse probability-of-treatment weighting (IPW), outcome regression (REG) and doubly robust (DR) estimation methods described in \citep{mcnealis2023doubly} to recover causal effects from the observed outcome, exposure and pretreatment covariates. The IPW estimator is consistent if the model for the exposure is correctly specified while consistency of the REG estimator will be achieved if the model for the outcome is correctly specified. The DR estimator, a bias-corrected regression estimator, is consistent if either the exposure or the outcome model is correctly specified. See \citep{mcnealis2023doubly} for additional details. Network interference estimators have been shown to perform well on synthetic computer-generated networks \citep{liu2016inverse, lee2021estimating, mcnealis2023doubly}, but simulation studies utilizing plasmodes or synthetic plasmodes are limited \citep{forastiere2021identification}. Moreover, the impact of unmeasured homophily confounding on interference estimators has not been assessed via a realistic simulation study. 

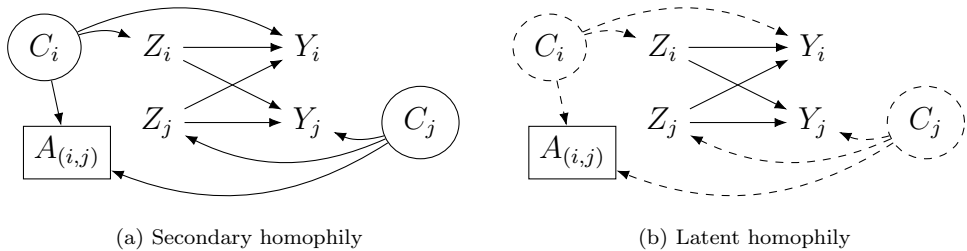
\begin{figure}[H]
\tikzset{
    -Latex,auto,node distance =1 cm and 1 cm,semithick,
    state/.style ={ellipse, draw, minimum width = 0.7 cm},
    point/.style = {circle, draw, inner sep=0.04cm,fill,node contents={}},
    bidirected/.style={Latex-Latex,dashed},
    el/.style = {inner sep=2pt, align=left, sloped}
}
\begin{subfigure}[b]{.45\linewidth}
 \begin{tikzpicture}
    \node (1) at (0,0) {$Z_{ i}$};
    \node (2) at (0, -1) {$Z_{ j}$};
    \node (3) at (2, 0) {$Y_{ i}$};
    \node (4) at (2, -1) {$Y_{ j}$};
     \node[state] (5) at (-1.5, 0) {$C_{i}$};
     \node[state] (6) at (3.5, -1) {$C_{j}$};
     \node[state, rectangle] (7) at (-1.2, -1.4) {$A_{(i,j)}$};
    \path (2) edge (3);
    \path (1) edge (3);
    \path (2) edge (4);
    \path (1) edge (4);
    \path (5) edge[bend left=20] (1);
    \path (5) edge[bend left=25] (3);
    \path (6) edge[bend left=25] (2);
    \path (6) edge[bend left=20] (4);
    \path (5) edge(7);
    \path (6) edge[bend left=27] (7);
\end{tikzpicture}
\caption{Secondary homophily}
\label{fig:secondaryhomophily}
\end{subfigure}
\quad
    \begin{subfigure}[b]{.45\linewidth}
           \begin{tikzpicture}
    \node (1) at (0,0) {$Z_{ i}$};
    \node (2) at (0, -1) {$Z_{ j}$};
    \node (3) at (2, 0) {$Y_{ i}$};
    \node (4) at (2, -1) {$Y_{ j}$};
     \node[state, dashed] (5) at (-1.5, 0) {$C_{i}$};
     \node[state, dashed] (6) at (3.5, -1) {$C_{j}$};
     \node[state, rectangle] (7) at (-1.2, -1.4) {$A_{(i,j)}$};
    \path (2) edge (3);
    \path (1) edge (3);
    \path (2) edge (4);
    \path (1) edge (4);
    \path[dashed] (5) edge[bend left=20] (1);
    \path[dashed] (5) edge[bend left=25] (3);
    \path[dashed] (6) edge[bend left=25] (2);
    \path[dashed] (6) edge[bend left=20] (4);
    \path[dashed] (5) edge(7);
    \path[dashed] (6) edge[bend left=27] (7);
\end{tikzpicture}
\caption{Latent homophily}
\label{fig:latenthomophily}
\end{subfigure}
\caption{Directed acyclic graphs displaying measured homophily confounding (Subfigure (a)) and unmeasured homophily confounding (Subfigure (b)) by a variable $C$ for two blocks $i$ and $j$. Pretreatment covariates $\bm{X}_{ i}$ and $\bm{X}_{ j}$ are excluded from the figure for simplicity.}
\label{fig:homophily}
        \end{figure}

We compared IPW, REG, and DR estimators based on correctly specified outcome and exposure models with estimators based on misspecified models due to the omission of a `regular' confounder (a confounder of the treatment-outcome relationship but which does not influence the formation of ties) or the omission of a homophilous confounder. Figure \ref{fig:homophily} shows an example of a variable $C$ that is not only linked to the exposure and outcome generating mechanisms but to the mechanism of tie formation as well. If unmeasured, $C$ cannot be conditioned upon, leading to an instance of latent homophily as shown in Figure \ref{fig:latenthomophily}. In our simulation design, latent homophily confounding can be induced by omitting the race variable from the models for the outcome and the exposure, since students with the same value for race had a higher probability of forming ties and race predicted both maternal education and student GPA. Alternatively, excluding the indicator of whether the father lives at home would only result in an instance of `regular' confounding, since this variable was not part of the ERGM used to generate the networks.

\begin{figure}
\includegraphics[scale=0.38]{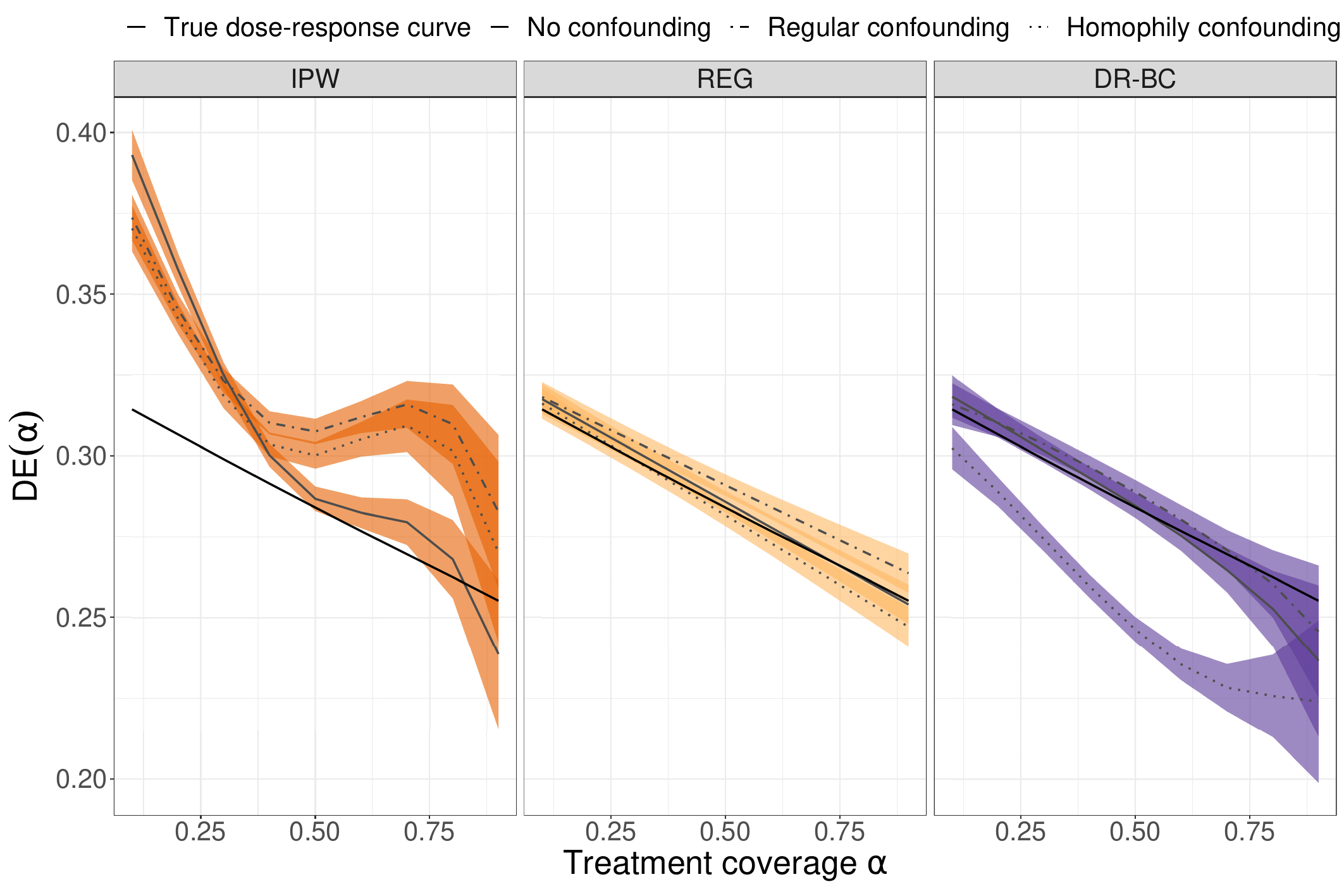}
\caption[Average estimates of the dose response curve $DE(\alpha)$ based on 500 simulation replicates under the first exposure generating scheme.]{IPW, REG, and DR-BC estimates of the dose response curve $DE(\alpha)$ under the first exposure generating scheme. The black solid lines represent the true dose-response curve, while the grey solid lines represent estimators assuming no confounding. The dashed and dotted lines represent estimators based on excluding a regular confounder and a homophilous confounder from the outcome and exposure models, respectively. The colored areas represents 95\% pointwise confidence intervals based on $S = 500$ simulation replicates.} 
    \label{fig:plot_first_scheme_paper3}
\end{figure}

Table \ref{tab:tab1_paper3} shows the bias and the mean squared error (MSE) of the IPW, REG, and DR-BC estimators of $DE(0.2)$, $DE(0.5)$, $DE(0.8)$, $IE(0.5, 0.2)$, $IE(0.8, 0.2)$, and $IE(0.8, 0.5)$ under different scenarios of confounding (no unmeasured confounding, regular confounding, and homophily confounding) for the first exposure generating scheme, in which the empirical bivariate associations between the exposure and selected pretreatment covariates were preserved. Figure \ref{fig:plot_first_scheme_paper3} displays the average of the 500 IPW, REG, and DR-BC estimators of the direct effect $DE(\alpha)$ for the first exposure generating scheme. In accordance with the model used to generate the potential outcomes, the dose-response curve $DE(\alpha)$ decreases with $\alpha$, while the indirect treatment effect $IE(\alpha, \alpha')$ increases for larger gaps between coverage $\alpha'$ and the reference coverage $\alpha$. The poor performance of the IPW estimator under the first exposure-generating scheme, can be explained by the fact that the posited treatment model did not reflect the true data generating process, as we used Model (\ref{eqn:exposuregeneration}) to estimate the propensity scores. The REG estimator performs better, with estimates closer to the true dose-response curve, especially under no unmeasured confounding. The REG and DR-BC estimators generally show similar performance, except for the scenario of homophily confounding, where DR-BC displays high bias. This is presumably because since both models are slightly off due to the confounding structure, the bias correction in the DR-BC method amplifies errors rather than mitigating them, resulting in higher bias. Two wrong models might be no better than one, as put by Kang and Schafer (2007) \cite{kang2007demystifying}. 

Table \ref{tab:tab2_paper3} shows the bias and the mean squared error (MSE) of the IPW, REG, and DR-BC estimators for the same estimands under the second exposure generating scheme, in which the exposure was generated according to Model (\ref{eqn:exposuregeneration}). Since the oracle model was used to construct the propensity scores, the IPW estimator shows better performance under this data generating scheme, with estimates being closer to the true estimands on average. This aside, the trends are consistent with the first exposure generating scheme, with higher bias and MSE observed when either the indicator that the father lives at home (regular confounding) or the student's race (homophily confounding) is excluded from the models. In summary, across both exposure generating schemes, IPW, REG, and DR-BC show increased bias and MSE under homophily confounding, with DR-BC being more affected.
        
\begin{table}[H]
\centering
\scriptsize
\caption{Bias and mean squared error (MSE) of the IPW, REG, and DR-BC estimators under the first exposure generating scheme.}
\begin{tabular}{c c c c c c c c c }
\toprule
    & &  & \multicolumn{2}{c}{IPW} & \multicolumn{2}{c}{REG} & \multicolumn{2}{c}{DR-BC} \\
    \cmidrule(l){4-5}
    \cmidrule(l){6-7}
    \cmidrule(l){8-9} 
  Scenario & Estimand &Value & $\text{Bias}$ & MSE & $\text{Bias}$ & MSE & $\text{Bias}$ & MSE \\
   \midrule
  No unmeasured  & $DE(0.2)$ & 0.307 & 0.051 & 0.006 & 0.003 & 0.002 & 0.003 & 0.003 \\ 
  confounding & $DE(0.5)$ & 0.284 & 0.003 & 0.002 & 0.002 & 0.001 & 0.001 & 0.002 \\ 
 & $DE(0.8)$ & 0.263 & 0.006 & 0.019 & -0.001 & 0.004 & -0.010 & 0.018 \\
 & $IE(0.5,0.2)$ & 0.120 & 0.002 & 0.002 & 0.001 & 0.001 & -0.001 & 0.002 \\ 
  & $IE(0.8,0.2)$ & 0.241 & 0.006 & 0.022 & 0.002 & 0.002 & -0.001 & 0.018 \\
  & $IE(0.8,0.5)$ & 0.120 & 0.004 & 0.018 & 0.001 & 0.001 & $<0.001$ & 0.014 \\ 
  \addlinespace
 Regular & $DE(0.2)$ & 0.307 & 0.039 & 0.004 & 0.005 & 0.002 & 0.004 & 0.002 \\
 confounding & $DE(0.5)$ & 0.284 & 0.023 & 0.003 & 0.007 & 0.001 & 0.005 & 0.002 \\ 
  & $DE(0.8)$ & 0.263 & 0.047 & 0.022 & 0.008 & 0.004 & -0.002 & 0.014 \\ 
  & $IE(0.5,0.2)$ & 0.120 & $<0.001$ & 0.002 & 0.001 & 0.001 & -0.001 & 0.002 \\
  & $IE(0.8,0.2)$ & 0.241 & $<0.001$ & 0.019 & 0.002 & 0.002 & -0.001 & 0.016 \\
  & $IE(0.8,0.5)$ & 0.120 & $<0.001$ & 0.015 & 0.001 & 0.001 & $<0.001$ & 0.013 \\
  \addlinespace
 Homophily & $DE(0.2)$ & 0.307 & 0.036 & 0.004 & 0.001 & 0.002 & -0.018 & 0.003 \\
 confounding & $DE(0.5)$ & 0.284 & 0.016 & 0.002 & -0.002 & 0.001 & -0.038 & 0.003 \\ 
  & $DE(0.8)$ & 0.263 & 0.039 & 0.028 & -0.007 & 0.004 & -0.037 & 0.022 \\ 
  & $IE(0.5,0.2)$ & 0.120 & 0.003 & 0.002 & 0.003 & 0.001 & 0.006 & 0.002 \\ 
  & $IE(0.8,0.2)$ & 0.241 & 0.006 & 0.021 & 0.007 & 0.002 & 0.013 & 0.020 \\ 
  & $IE(0.8,0.5)$ & 0.120 & 0.004 & 0.016 & 0.004 & 0.001 & 0.007 & 0.015 \\ 
  \bottomrule
\end{tabular}
\label{tab:tab1_paper3}
\end{table}

\begin{table}[H]
\centering
\scriptsize
\caption{Bias and mean squared error (MSE) of the IPW, REG, and DR-BC estimators under the second exposure generating scheme.}
\begin{tabular}{c c c c c c c c c }
\toprule
    & &  & \multicolumn{2}{c}{IPW} & \multicolumn{2}{c}{REG} & \multicolumn{2}{c}{DR-BC} \\
    \cmidrule(l){4-5}
    \cmidrule(l){6-7}
    \cmidrule(l){8-9} 
  Scenario & Estimand &Value & $\text{Bias}$ & MSE & $\text{Bias}$ & MSE & $\text{Bias}$ & MSE \\
   \midrule
  No unmeasured  & $DE(0.2)$ & 0.307 & 0.006 & 0.003 & -0.001 & 0.002 & -0.003 & 0.003 \\ 
  confounding & $DE(0.5)$ & 0.284 & 0.008 & 0.003 & -0.004 & 0.002 & -0.004 & 0.003 \\ 
 & $DE(0.8)$ & 0.263 & -0.013 & 0.05 & -0.008 & 0.005 & -0.014 & 0.038 \\ 
 & $IE(0.5,0.2)$ & 0.120 & 0.002 & 0.002 & 0.001 & 0.001 & -0.001 & 0.002 \\
  & $IE(0.8,0.2)$ & 0.241 & 0.006 & 0.022 & 0.002 & 0.002 & -0.001 & 0.018 \\ 
  & $IE(0.8,0.5)$  & 0.120 & 0.004 & 0.018 & 0.001 & 0.001 & $<0.001$ & 0.014 \\ 
  \addlinespace
 Regular & $DE(0.2)$ & 0.307 & 0.004 & 0.003 & 0.003 & 0.002 & 0.002 & 0.003 \\
 confounding & $DE(0.5)$ & 0.284 & 0.001 & 0.003 & 0.001 & 0.002 & $<0.001$ & 0.003 \\ 
  & $DE(0.8)$ & 0.263 & -0.020 & 0.048 & -0.003 & 0.005 & -0.011 & 0.033 \\ 
  & $IE(0.5,0.2)$ & 0.120 & $<0.001$ & 0.002 & 0.001 & 0.001 & -0.001 & 0.002 \\ 
  & $IE(0.8,0.2)$ & 0.241 & $<0.001$  & 0.019 & 0.002 & 0.002 & -0.001 & 0.016 \\ 
  & $IE(0.8,0.5)$ & 0.120 & $<0.001$ & 0.015 & 0.001 & 0.001 & $<0.001$ & 0.013 \\ 
  \addlinespace
 Homophily & $DE(0.2)$  & 0.307 & 0.007 & 0.003 & 0.009 & 0.002 & 0.009 & 0.003 \\
 confounding & $DE(0.5)$ & 0.284 & 0.005 & 0.003 & 0.006 & 0.002 & 0.007 & 0.002 \\
  & $DE(0.8)$ & 0.263 & -0.016 & 0.049 & 0.002 & 0.005 & -0.007 & 0.037 \\ 
  & $IE(0.5,0.2)$ & 0.120 & 0.003 & 0.002 & 0.003 & 0.001 & 0.006 & 0.002 \\ 
  & $IE(0.8,0.2)$ & 0.241 & 0.006 & 0.021 & 0.007 & 0.002 & 0.013 & 0.020 \\ 
  & $IE(0.8,0.5)$ & 0.120 & 0.004 & 0.016 & 0.004 & 0.001 & 0.007 & 0.015 \\
  \bottomrule
\end{tabular}
\label{tab:tab2_paper3}
\end{table}

\section{Discussion}
\label{section:discussion}

In this work, we have outlined steps for creating simulated point-treatment network-based observational studies from private network data to evaluate the performance of analytic strategies of interest to the researcher. We demonstrated our simulation framework with a study of spillover effects of home environment on adolescent academic performance leveraging Add Health data and compared the simulated data to the observed data. The simulation framework produced datasets closely mirroring the complex data structure observed in five selected schools in the Add Health study. Using the simulated datasets, we provided an example in which we used this framework to evaluate statistical methods in the presence of latent homophily, which had not previously be assessed in realistic simulated data. Our findings indicate that latent homophily threatens the inference for interference effects, corroborating previous theoretical and empirical results established from toy models and computer-generated network data \citep{shalizi2011homophily, mcfowland2023estimating, mcnealis2023doubly}. Further, our simulation results are completely reproducible, having provided all model estimates and sufficient statistics required to generate synthetic datasets.

As with all proposals, ours also comes with some limitations. In our application, graphs and outcomes are produced using a limited number of covariates, whereas in reality, friendship formation and academic performance are affected by a much broader array of factors, both measured and unmeasured. Given that the nodal attributes were private, we had to synthesize covariate data using a Gaussian-copula method for categorical data which limited the set of covariates that could be included in the study. An investigator need not be limited to the specific technique we used. With other patterns of nodal attributes, such as mixtures of continuous and discrete variables, it may be well advised to investigate the other synthetic data generators discussed in this paper. In addition, we specified a single ERGM for the five selected schools, which constrained the terms that could be included for consideration in the model since the covariate distribution (e.g., the distribution of the race variable) varied considerably across different schools. A workaround would have been to tailor the model specification to each school. 

For the purpose of analyzing interference effects, we restricted ourselves to five schools of moderate size and augmented our sample size (number of clusters) for the implementation of causal inference methods by replicating the schools in the simulated datasets. As a result, our final simulated datasets are not as realistic as they could be. At the cost of greater computational time, it would have been possible to apply the simulation framework to a wider variety of schools in Add Health, including the largest school of 2,209 nodes, as was demonstrated by Hunter et al.~(2008) who successfully fitted an ERGM on this school \citep{hunter2008goodness}. However, in this specific case, larger schools in the sample would have made the evaluation of causal inference methods by simulation prohibitively computationally expensive without providing any real gain in insights to our main proposal, which aimed to provide a plasmode approach to simulating network data.

There are several potential avenues for future work. Within a cross-sectional or point-treatment setting, an increasingly popular alternative to the ERGM for network modeling and simulation is the stochastic block model, which assumes that the nodes of the network are partitioned into latent blocks, and that the probability of a tie between two nodes depends on the blocks to which the nodes belong \citep{holland1983stochastic, kolaczyk2009}. This method presents the advantage of providing insights into the latent structure of a network. Recently, Kitamura and Laage (2024) introduced a stochastic block model which incorporates covariates \citep{kitamura2024estimating}. Another natural direction would be to extend the simulation framework to longitudinal settings. For instance, an analogous simulation framework could be devised for dynamic networks, through the use of temporal ERGMs \citep{leifeld2018temporal} or stochastic actor-oriented models, which allow to model the co-evolution of nodal attributes and the graph \citep{snijders2001statistical, snijders2005models}.

\section*{Acknowledgments}

Vanessa McNealis is supported by doctoral fellowships from Natural Sciences and Engineering Research Council of Canada (NSERC) and the Fonds de Recherche du Québec (FRQ) - Nature et Technologie. Erica E. M. Moodie acknowledges support from an NSERC Discovery Grant. Erica E. M. Moodie is a Canada Research Chair (Tier 1) in Statistical Methods for Precision Medicine and acknowledges the support of a Chercheur-boursier de mérite career award from the FRQ - Santé. 

This research was enabled in part by support provided by the Digital Research Alliance of Canada (\url{https://alliancecan.ca/en}). The case study in this paper uses data from Add Health, funded by grant P01 HD31921 (Harris) from the Eunice Kennedy Shriver National Institute of Child Health and Human Development (NICHD), with cooperative funding from 23 other federal agencies and foundations. Add Health is currently directed by Robert A. Hummer and funded by the National Institute on Aging cooperative agreements U01 AG071448 (Hummer) and U01AG071450 (Aiello and Hummer) at the University of North Carolina at Chapel Hill. Add Health was designed by J. Richard Udry, Peter S. Bearman, and Kathleen Mullan Harris at the University of North Carolina at Chapel Hill.

\FloatBarrier

 \bibliographystyle{elsarticle-harv} 
 \bibliography{elsarticle-template-num}

 \newpage
\appendix
\setcounter{figure}{0} 
\setcounter{table}{0} 
\section{Variable definitions in the Add Health study}
\label{appendix:variables}
\begin{table}[H]
\centering
\footnotesize
\caption{Definitions of outcome, treatment, and covariates}
\begin{tabular}{lll}
\toprule
Variable & Definition & Support \\
\midrule
$Y$ & Standardized Grade Point Average & $[-2.30, 1.50]$\\
$Z$ & Mother holds a 4-year college degree & \{No, Yes\} \\
$X_1$ & Male student & \{No, Yes\} \\
$X_2$ & Grade & \{7, 8, 9, 10, 11, 12\} \\
$X_3$ & Race & \{White, Black, Asian, Other\}\\
$X_4$ & Father lives at home & \{No, Yes\}\\
$X_5$ & Screen time & \{$\leq$ 2 h per day, $>2$ h per day\}\\
$X_6$ & Motivation & \{High, Moderate, Low\} \\
$X_7$ & Sense of belonging & \{No, Yes\}\\
$X_8$ & Physically fit & \{No, Yes\}\\
\bottomrule
\end{tabular}
\end{table}
\newpage
\section{Network characteristics by school}
\label{appendix:networkcharacteristics}

Table \ref{tab:networkcharacteristics} displays network characteristics for each of the five schools. We provide an example of interpretation of the graph theoretical properties in the context of the Add Health variables. For instance, the average degree (i.e., average number of named friends per participant) in School 003, defined as $N^{-1} \sum_{i \in \mathcal{N}} d_{i}$, is equal to 6.202 and the standard deviation of the corresponding degree distribution is 3.938. The edge density, defined as the ratio of the number of edges to the number of possible edges, has value 0.063 in School 003, indicating a weakly connected friendship network. To define transitivity, we first need to define a triangle, which is a complete subgraph of order 3, and a connected triple, which is a subgraph of three nodes connected by two edges \citep{kolaczyk2009}. As shown in Table \ref{tab:networkcharacteristics}, the transitivity or clustering coefficient, defined as the ratio of the count of \emph{triangles} to the count of connected triples, is 0.255, which quantifies the extent to which edges are clustered in the graph \citep{kolaczyk2009}. Finally, assortativity, which can be seen as a correlation coefficient, measures the extent to which nodes with similar values of a nodal attribute tend to form ties in the network. The assortativity coefficient based on school grade ($\in \{7,8,9,10,11,12\}$) is defined as $$r = \frac{\sum_{i=7}^{12} f_{ii} - \sum_{i=7}^{12} f_{i+}f_{+i}}{1-\sum_{i=7}^{12} f_{i+}f_{+i}},$$ where $f_{ij}$ is the fraction of edges in $G$ that join a node in the $i$-th category with a node in the $j$-th category, $f_{i+} = \sum_{j=0}^1 f_{ij}$ and $f_{+i} = \sum_{j=0}^1 f_{ji}$ \citep{igraph, kolaczyk2009}. The support of $r$ is the interval $[-1,1]$ The assortativity in School 003 is 0.682, indicating that friends tend to belong to the same school grade. 

\begin{table}[H]
\centering
\caption{Characterization of the Add Health subnetwork of five schools (School 003, School 028, School 106, School 122, and School 173) after exclusion of isolates}
    \small
    \begin{tabular}{l l c c }
    \toprule
       School 003  & Nodes ($N_1$) & \multicolumn{2}{c}{99} \\
         & Edges & \multicolumn{2}{c}{307}
          \\
         & Average Degree (SD) & \multicolumn{2}{c}{6.20 (3.94)}\\
         & Edge density & \multicolumn{2}{c}{0.06}\\
         & Transitivity & \multicolumn{2}{c}{0.17} \\
         & Assortativity\textsuperscript{a}& \multicolumn{2}{c}{0.68} \\
         \midrule
         School 028  & Nodes ($N_2$) & \multicolumn{2}{c}{88} \\
         & Edges & \multicolumn{2}{c}{201}
          \\
         & Average Degree (SD) & \multicolumn{2}{c}{4.57 (2.48)}\\
         & Edge density & \multicolumn{2}{c}{0.05}\\
         & Transitivity & \multicolumn{2}{c}{0.28} \\
         & Assortativity\textsuperscript{a}& \multicolumn{2}{c}{0.64} \\
        \midrule
         School 106  & Nodes ($N_3$) & \multicolumn{2}{c}{71} \\
         & Edges & \multicolumn{2}{c}{262}
          \\
         & Average Degree (SD) & \multicolumn{2}{c}{7.38 (3.54)}\\
         & Edge density & \multicolumn{2}{c}{0.11}\\
         & Transitivity & \multicolumn{2}{c}{0.30} \\
         & Assortativity\textsuperscript{a}& \multicolumn{2}{c}{0.79} \\
         \midrule
         School 122  & Nodes ($N_4$) & \multicolumn{2}{c}{101} \\
         & Edges & \multicolumn{2}{c}{385}
          \\
         & Average Degree (SD) & \multicolumn{2}{c}{7.62 (3.94)}\\
         & Edge density & \multicolumn{2}{c}{0.08}\\
         & Transitivity & \multicolumn{2}{c}{0.34} \\
         & Assortativity\textsuperscript{a}& \multicolumn{2}{c}{0.80} \\
         \midrule
         School 173  & Nodes ($N_5$) & \multicolumn{2}{c}{86} \\
         & Edges & \multicolumn{2}{c}{111}
          \\
         & Average Degree (SD) & \multicolumn{2}{c}{2.58 (1.73)}\\
         & Edge density & \multicolumn{2}{c}{0.03}\\
         & Transitivity & \multicolumn{2}{c}{0.27} \\
         & Assortativity\textsuperscript{a}& \multicolumn{2}{c}{0.09} \\
         \bottomrule
    \end{tabular}
\begin{tablenotes}
\item \textsuperscript{a} The assortativity coefficient is based on school grade.
\end{tablenotes}
    \label{tab:networkcharacteristics}
\end{table}

\section{Sufficient statistics and model estimates for School 003}
\label{appendix:school003}
\begin{figure}[H]
\includegraphics[scale=0.54]{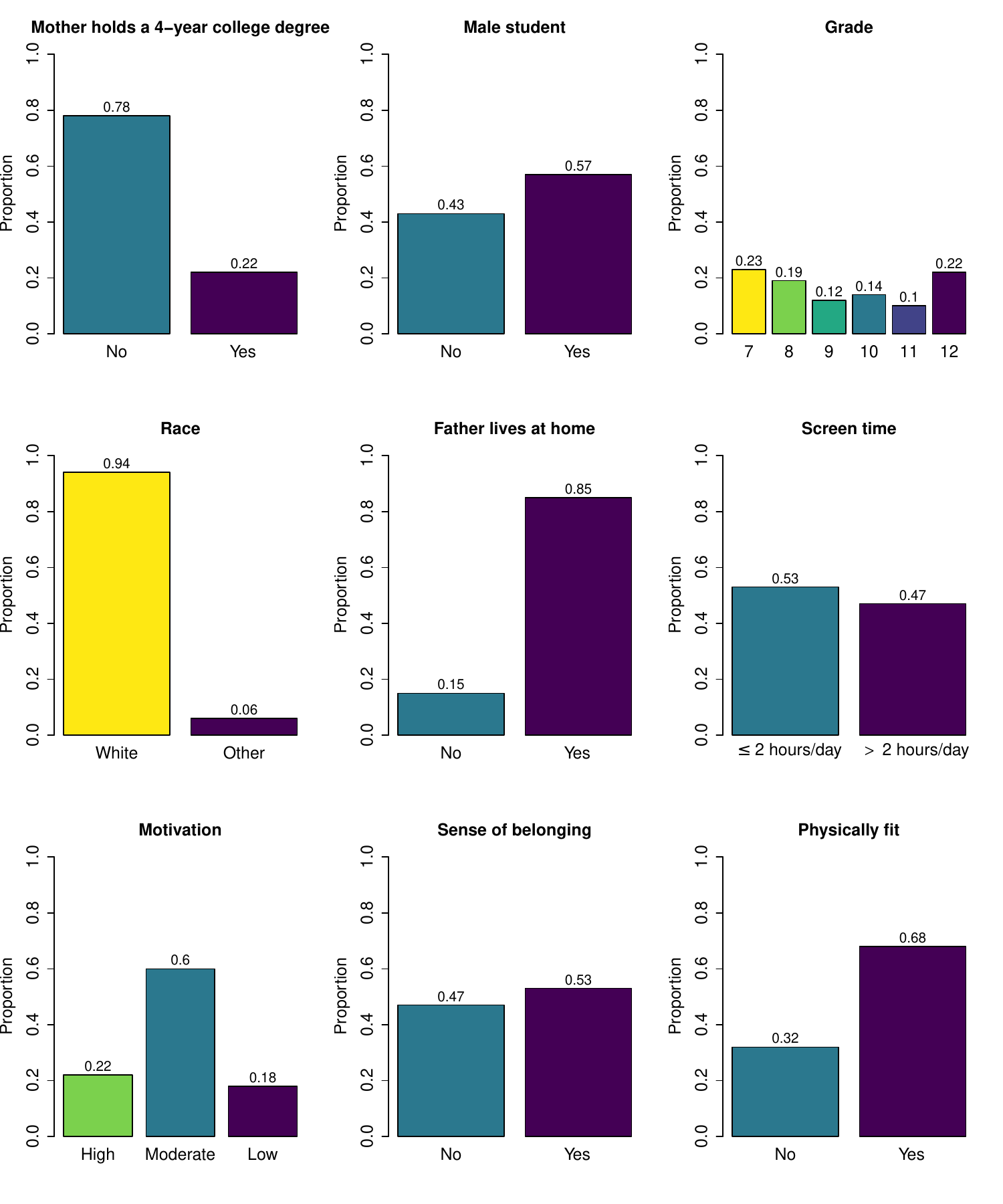}
\caption{Observed marginal distributions for the exposure and selected categorical pretreatment covariates in School 003.}
\label{fig:marginals003}
\end{figure}

\begin{table}[H]
\caption{Estimated Spearman's rank correlation matrix for the exposure and selected categorical pretreatment covariates in School 003. See \ref{appendix:variables} for variable names.}
\label{tab:spearman003}
\centering
\begin{tabular}{l lllllllll}
  & $Z$ & $X_1$ & $X_2$ & $X_3$ & $X_4$ & $X_5$ & $X_6$ & $X_7$ & $X_8$\\
  \hline
$Z$ & \cellcolor[HTML]{1A1AE6}{ 1.00} & \cellcolor[HTML]{7777F0}{ 0.17} & \cellcolor[HTML]{A6A6F6}{-0.22} & \cellcolor[HTML]{7777F0}{ 0.07} & \cellcolor[HTML]{9797F4}{-0.11} & \cellcolor[HTML]{9797F4}{-0.12} & \cellcolor[HTML]{A6A6F6}{-0.20} & \cellcolor[HTML]{7777F0}{ 0.07} & \cellcolor[HTML]{8787F2}{ 0.06} \\ 
$X_1$ &  \cellcolor[HTML]{7777F0}{ 0.17} & \cellcolor[HTML]{1A1AE6}{ 1.00} & \cellcolor[HTML]{9797F4}{-0.12} & \cellcolor[HTML]{8787F2}{-0.03} & \cellcolor[HTML]{7777F0}{ 0.08} & \cellcolor[HTML]{8787F2}{-0.02} & \cellcolor[HTML]{7777F0}{ 0.10} & \cellcolor[HTML]{7777F0}{ 0.19} & \cellcolor[HTML]{6868EE}{ 0.22} \\ 
 $X_2$ & \cellcolor[HTML]{A6A6F6}{-0.22} & \cellcolor[HTML]{9797F4}{-0.12} & \cellcolor[HTML]{1A1AE6}{ 1.00} & \cellcolor[HTML]{9797F4}{-0.07} & \cellcolor[HTML]{9797F4}{-0.15} & \cellcolor[HTML]{8787F2}{-0.02} & \cellcolor[HTML]{7777F0}{ 0.18} & \cellcolor[HTML]{A6A6F6}{-0.24} & \cellcolor[HTML]{8787F2}{-0.01} \\ 
 $X_3$ & \cellcolor[HTML]{7777F0}{ 0.07} & \cellcolor[HTML]{8787F2}{-0.03} & \cellcolor[HTML]{9797F4}{-0.07} & \cellcolor[HTML]{1A1AE6}{ 1.00} & \cellcolor[HTML]{8787F2}{-0.01} & \cellcolor[HTML]{8787F2}{ 0.01} & \cellcolor[HTML]{8787F2}{ 0.02} & \cellcolor[HTML]{9797F4}{-0.10} & \cellcolor[HTML]{8787F2}{-0.01} \\ 
 $X_4$ & \cellcolor[HTML]{9797F4}{-0.11} & \cellcolor[HTML]{7777F0}{ 0.08} & \cellcolor[HTML]{9797F4}{-0.15} & \cellcolor[HTML]{8787F2}{-0.01} & \cellcolor[HTML]{1A1AE6}{ 1.00} & \cellcolor[HTML]{9797F4}{-0.11} & \cellcolor[HTML]{9797F4}{-0.07} & \cellcolor[HTML]{8787F2}{-0.01} & \cellcolor[HTML]{7777F0}{ 0.19} \\ 
  $X_5$ &\cellcolor[HTML]{9797F4}{-0.12} & \cellcolor[HTML]{8787F2}{-0.02} & \cellcolor[HTML]{8787F2}{-0.02} & \cellcolor[HTML]{8787F2}{ 0.01} & \cellcolor[HTML]{9797F4}{-0.11} & \cellcolor[HTML]{1A1AE6}{ 1.00} & \cellcolor[HTML]{8787F2}{-0.04} & \cellcolor[HTML]{8787F2}{-0.03} & \cellcolor[HTML]{8787F2}{-0.03} \\ 
 $X_6$ & \cellcolor[HTML]{A6A6F6}{-0.20} & \cellcolor[HTML]{7777F0}{ 0.10} & \cellcolor[HTML]{7777F0}{ 0.18} & \cellcolor[HTML]{8787F2}{ 0.02} & \cellcolor[HTML]{9797F4}{-0.07} & \cellcolor[HTML]{8787F2}{-0.04} & \cellcolor[HTML]{1A1AE6}{ 1.00} & \cellcolor[HTML]{9797F4}{-0.12} & \cellcolor[HTML]{9797F4}{-0.08} \\ 
 $X_7$ &  \cellcolor[HTML]{7777F0}{ 0.07} & \cellcolor[HTML]{7777F0}{ 0.19} & \cellcolor[HTML]{A6A6F6}{-0.24} & \cellcolor[HTML]{9797F4}{-0.10} & \cellcolor[HTML]{8787F2}{-0.01} & \cellcolor[HTML]{8787F2}{-0.03} & \cellcolor[HTML]{9797F4}{-0.12} & \cellcolor[HTML]{1A1AE6}{ 1.00} & \cellcolor[HTML]{6868EE}{ 0.25} \\ 
 $X_8$ & \cellcolor[HTML]{8787F2}{ 0.06} & \cellcolor[HTML]{6868EE}{ 0.22} & \cellcolor[HTML]{8787F2}{-0.01} & \cellcolor[HTML]{8787F2}{-0.01} & \cellcolor[HTML]{7777F0}{ 0.19} & \cellcolor[HTML]{8787F2}{-0.03} & \cellcolor[HTML]{9797F4}{-0.08} & \cellcolor[HTML]{6868EE}{ 0.25} & \cellcolor[HTML]{1A1AE6}{ 1.00} \\ 
   \hline
\end{tabular}
\end{table}

\begin{table}[H]
\caption{Coefficient estimates for the ERGM fitted to School 003}
\label{tab:ergm003}
\centering

\begin{tabular}{lc}
  \toprule
 Term   &\multicolumn{1}{c}{Coefficient estimate} \\
  \midrule
Edges & -9.7001 \\
GWESP ($\lambda = 1$) & 0.6598 \\
GWD ($\lambda = 1$) & 0.6349 \\
NF Male & 0.2254  \\
NF Race (White) & 6.2567  \\
NF Grade (8) & 0.0226  \\
NF Grade (9) & -0.0804 \\
NF Grade (10) & 0.0237 \\
NF Grade (11) & 0.1957  \\
NF Grade (12) & -0.1081 \\
UH Male & 0.4385 \\
UH Race & -6.2688\\
UH Grade & 0.5269 \\
AD Grade & -0.7336 \\
\bottomrule
\end{tabular}
\centering
\scriptsize
\begin{tablenotes}
    \item[1] Abbreviations: GWESP = Geometrically Weighted Edgewise Shared Partner; GWD = Geometrically Weighted Degree; NF = Node Factor; UH = Uniform Homophily; AD = Absolute Difference.
    \item[2] Baseline levels for NF Race and NF Grade are Other and 7, respectively.
  \end{tablenotes}
\end{table}

\begin{figure}[H]
\includegraphics[scale=0.45]{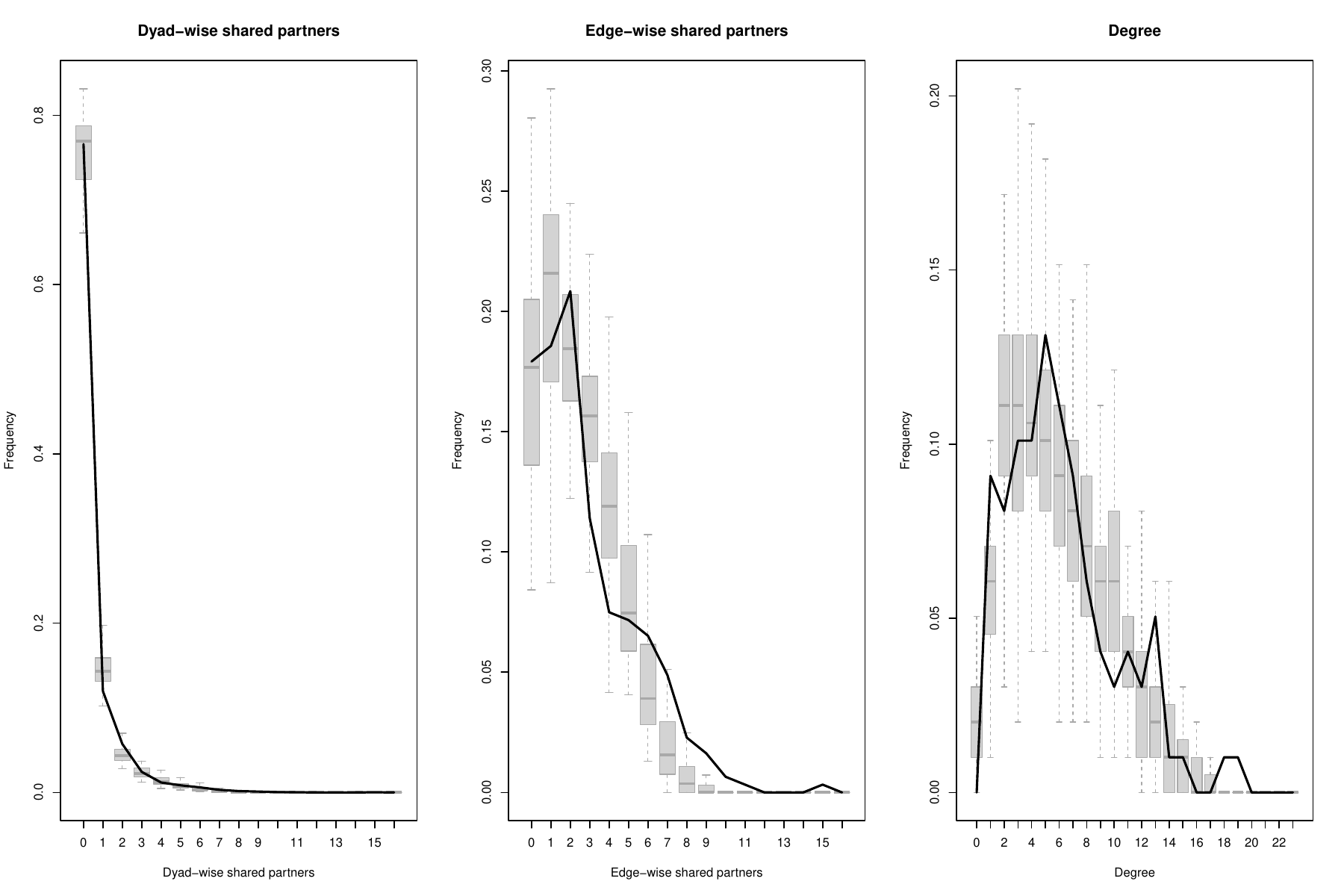}
\caption{Goodness-of-fit assessment of the ERGM for School 003 based on 100 simulated networks. The black solid line represents the observed distribution, whereas the boxplots display the distributions of the network statistics obtained across the 100 simulated networks.}
\label{fig:gof003}
\end{figure}

\section{Sufficient statistics and model estimates for School 028}
\label{appendix:school028}
\begin{figure}[H]
\includegraphics[scale=0.54]{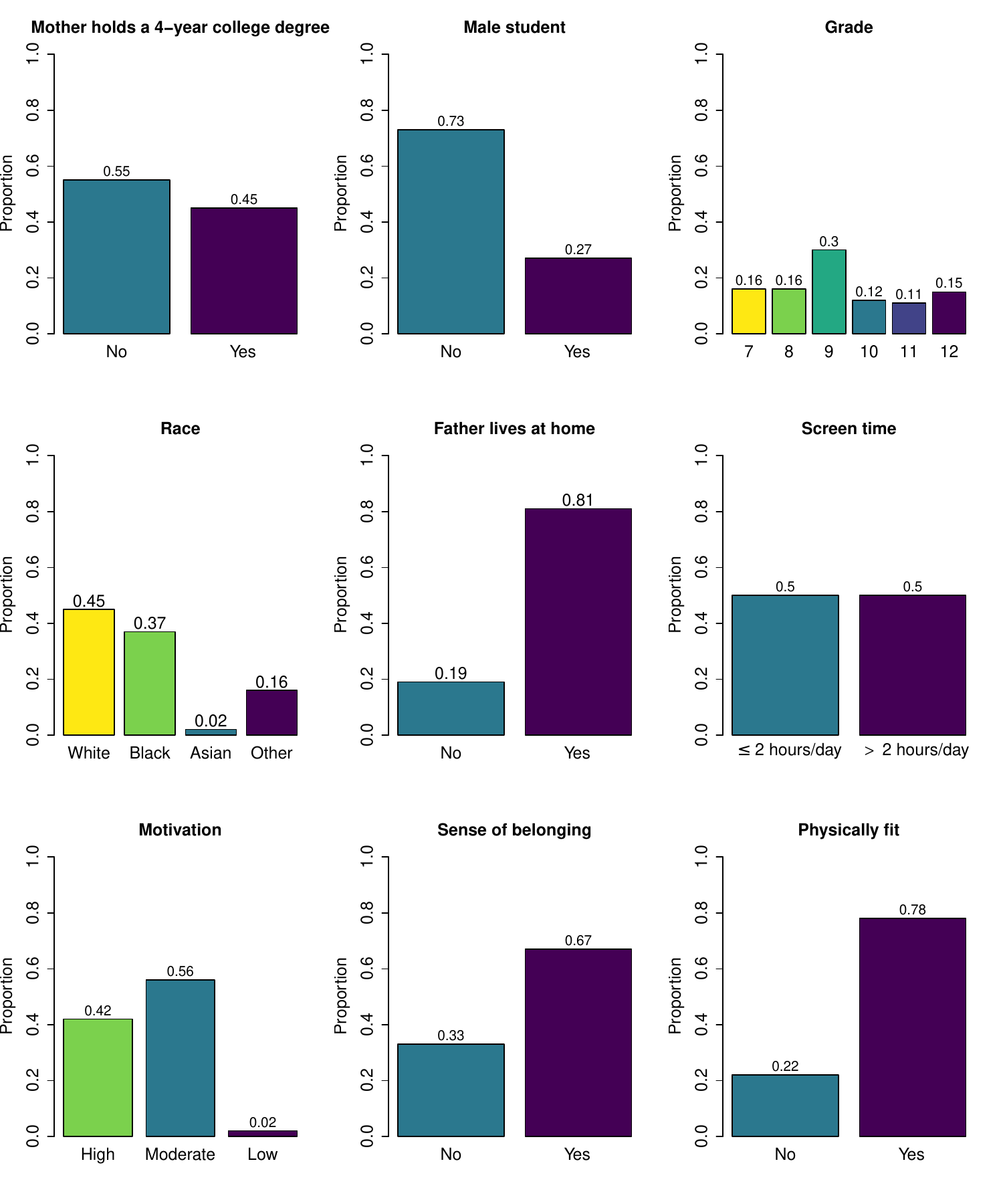}
\caption{Observed marginal distributions for the exposure and selected categorical pretreatment covariates in School 028.}
\label{fig:marginals028}
\end{figure}

\begin{table}[H]
\caption{Estimated Spearman's rank correlation matrix for the exposure and selected categorical attributes in School 028. See \ref{appendix:variables} for variable names.}
\label{tab:spearman028}
\centering
\begin{tabular}{l lllllllll}
  & $Z$ & $X_1$ & $X_2$ & $X_3$ & $X_4$ & $X_5$ & $X_6$ & $X_7$ & $X_8$\\
  \hline
$Z$ & \cellcolor[HTML]{1A1AE6}{ 1.00} & \cellcolor[HTML]{7777F0}{ 0.11} & \cellcolor[HTML]{A6A6F6}{-0.28} & \cellcolor[HTML]{8787F2}{-0.01} & \cellcolor[HTML]{6868EE}{ 0.33} & \cellcolor[HTML]{7777F0}{ 0.09} & \cellcolor[HTML]{9797F4}{-0.08} & \cellcolor[HTML]{8787F2}{-0.04} & \cellcolor[HTML]{7777F0}{ 0.09} \\ 
 $X_1$ &  \cellcolor[HTML]{7777F0}{ 0.11} & \cellcolor[HTML]{1A1AE6}{ 1.00} & \cellcolor[HTML]{8787F2}{ 0.00} & \cellcolor[HTML]{9797F4}{-0.11} & \cellcolor[HTML]{7777F0}{ 0.11} & \cellcolor[HTML]{9797F4}{-0.10} & \cellcolor[HTML]{7777F0}{ 0.14} & \cellcolor[HTML]{8787F2}{-0.06} & \cellcolor[HTML]{7777F0}{ 0.14} \\ 
  $X_2$ &  \cellcolor[HTML]{A6A6F6}{-0.28} & \cellcolor[HTML]{8787F2}{ 0.00} & \cellcolor[HTML]{1A1AE6}{ 1.00} & \cellcolor[HTML]{8787F2}{-0.01} & \cellcolor[HTML]{9797F4}{-0.15} & \cellcolor[HTML]{9797F4}{-0.08} & \cellcolor[HTML]{6868EE}{ 0.20} & \cellcolor[HTML]{9797F4}{-0.11} & \cellcolor[HTML]{A6A6F6}{-0.21} \\ 
 $X_3$ &  \cellcolor[HTML]{8787F2}{-0.01} & \cellcolor[HTML]{9797F4}{-0.11} & \cellcolor[HTML]{8787F2}{-0.01} & \cellcolor[HTML]{1A1AE6}{ 1.00} & \cellcolor[HTML]{8787F2}{-0.05} & \cellcolor[HTML]{5858ED}{ 0.41} & \cellcolor[HTML]{8787F2}{ 0.00} & \cellcolor[HTML]{9797F4}{-0.12} & \cellcolor[HTML]{8787F2}{-0.05} \\ 
 $X_4$ & \cellcolor[HTML]{6868EE}{ 0.33} & \cellcolor[HTML]{7777F0}{ 0.11} & \cellcolor[HTML]{9797F4}{-0.15} & \cellcolor[HTML]{8787F2}{-0.05} & \cellcolor[HTML]{1A1AE6}{ 1.00} & \cellcolor[HTML]{7777F0}{ 0.09} & \cellcolor[HTML]{8787F2}{ 0.00} & \cellcolor[HTML]{7777F0}{ 0.09} & \cellcolor[HTML]{7777F0}{ 0.09} \\ 
  $X_5$ &  \cellcolor[HTML]{7777F0}{ 0.09} & \cellcolor[HTML]{9797F4}{-0.10} & \cellcolor[HTML]{9797F4}{-0.08} & \cellcolor[HTML]{5858ED}{ 0.41} & \cellcolor[HTML]{7777F0}{ 0.09} & \cellcolor[HTML]{1A1AE6}{ 1.00} & \cellcolor[HTML]{7777F0}{ 0.09} & \cellcolor[HTML]{8787F2}{-0.02} & \cellcolor[HTML]{8787F2}{-0.03} \\ 
  $X_6$ & \cellcolor[HTML]{9797F4}{-0.08} & \cellcolor[HTML]{7777F0}{ 0.14} & \cellcolor[HTML]{6868EE}{ 0.20} & \cellcolor[HTML]{8787F2}{ 0.00} & \cellcolor[HTML]{8787F2}{ 0.00} & \cellcolor[HTML]{7777F0}{ 0.09} & \cellcolor[HTML]{1A1AE6}{ 1.00} & \cellcolor[HTML]{8787F2}{-0.02} & \cellcolor[HTML]{9797F4}{-0.07} \\ 
   $X_7$ &\cellcolor[HTML]{8787F2}{-0.04} & \cellcolor[HTML]{8787F2}{-0.06} & \cellcolor[HTML]{9797F4}{-0.11} & \cellcolor[HTML]{9797F4}{-0.12} & \cellcolor[HTML]{7777F0}{ 0.09} & \cellcolor[HTML]{8787F2}{-0.02} & \cellcolor[HTML]{8787F2}{-0.02} & \cellcolor[HTML]{1A1AE6}{ 1.00} & \cellcolor[HTML]{8787F2}{ 0.04} \\ 
$X_8$ &  \cellcolor[HTML]{7777F0}{ 0.09} & \cellcolor[HTML]{7777F0}{ 0.14} & \cellcolor[HTML]{A6A6F6}{-0.21} & \cellcolor[HTML]{8787F2}{-0.05} & \cellcolor[HTML]{7777F0}{ 0.09} & \cellcolor[HTML]{8787F2}{-0.03} & \cellcolor[HTML]{9797F4}{-0.07} & \cellcolor[HTML]{8787F2}{ 0.04} & \cellcolor[HTML]{1A1AE6}{ 1.00} \\  
   \hline
\end{tabular}
\end{table}

\begin{table}[H]
\caption{Coefficient estimates for the ERGM fitted to School 028}
\label{tab:ergm028}
\centering

\begin{tabular}{lc}
  \toprule
 Term   &\multicolumn{1}{c}{Coefficient estimate} \\
  \midrule
Edges & -3.7311 \\
GWESP ($\lambda = 1$) & 0.5572 \\
GWD ($\lambda = 1$) & 0.8890 \\
NF Male & 0.0830  \\
NF Race (Black) & -0.2430  \\
NF Race (Other) & 0.0662 \\
NF Race (White) & -0.2438 \\
NF Grade (8) & -0.2394  \\
NF Grade (9) & -0.3191 \\
NF Grade (10) & -0.1067 \\
NF Grade (11) & 0.1431 \\
NF Grade (12) & 0.0376 \\
UH Male & -0.0089 \\
UH Race & 1.2674\\
UH Grade & 0.9988 \\
AD Grade &-0.5843 \\
\bottomrule
\end{tabular}
\scriptsize
\begin{tablenotes}
    \item[1] Abbreviations: GWESP = Geometrically Weighted Edgewise Shared Partner; GWD = Geometrically Weighted Degree; NF = Node Factor; UH = Uniform Homophily.
    \item[2] Baseline levels for NF Race and NF Grade are White and 7, respectively.
  \end{tablenotes}
\end{table}

\begin{figure}[H]
\includegraphics[scale=0.45]{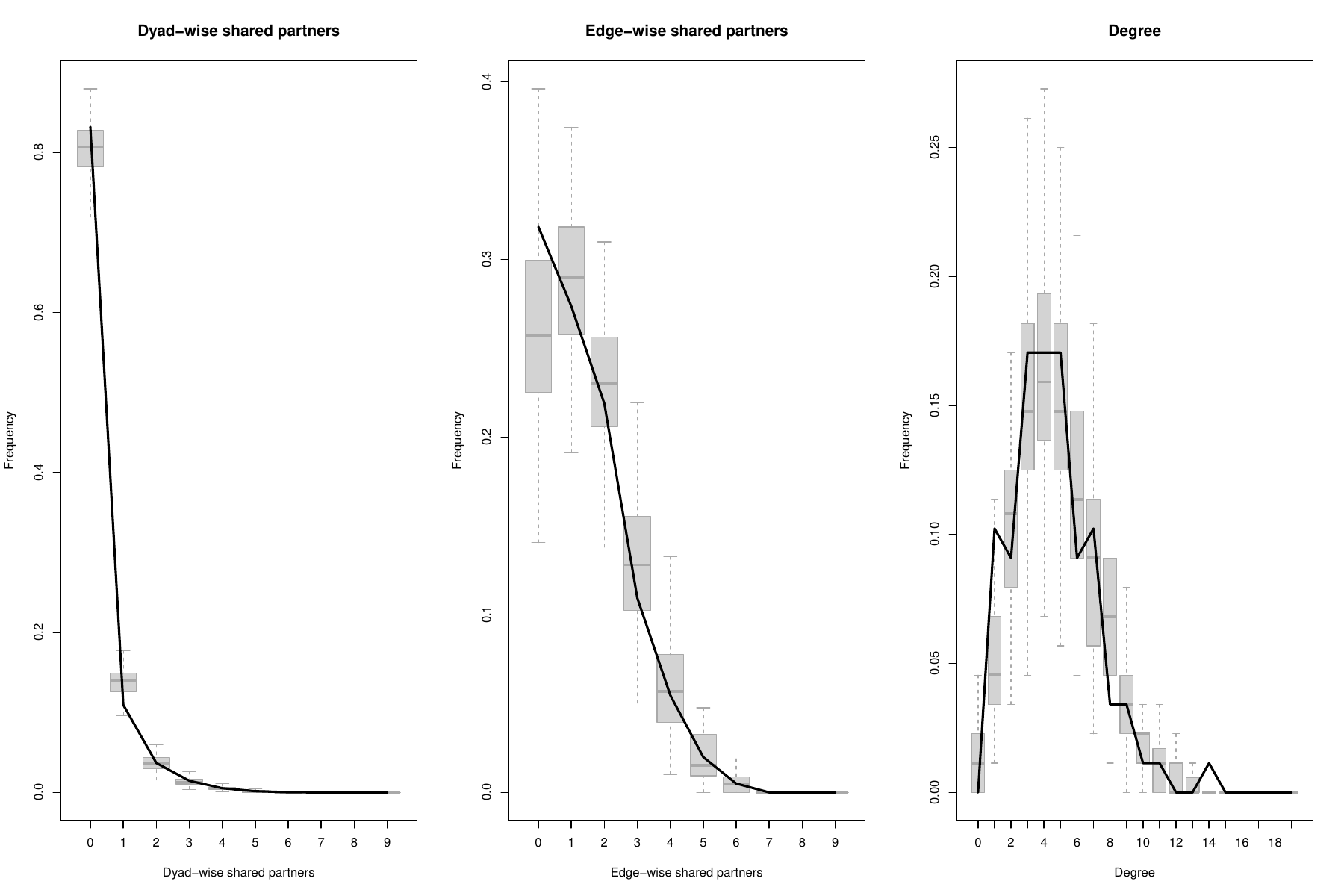}
\caption{Goodness-of-fit assessment of the ERGM for School 028 based on 100 simulated networks. The black solid line represents the observed distribution, whereas the boxplots display the distributions of the network statistics obtained across the 100 simulated networks.}
\label{fig:gof028}
\end{figure}

\section{Sufficient statistics and model estimates for School 106}
\label{appendix:school106}
\begin{figure}[H]
\includegraphics[scale=0.54]{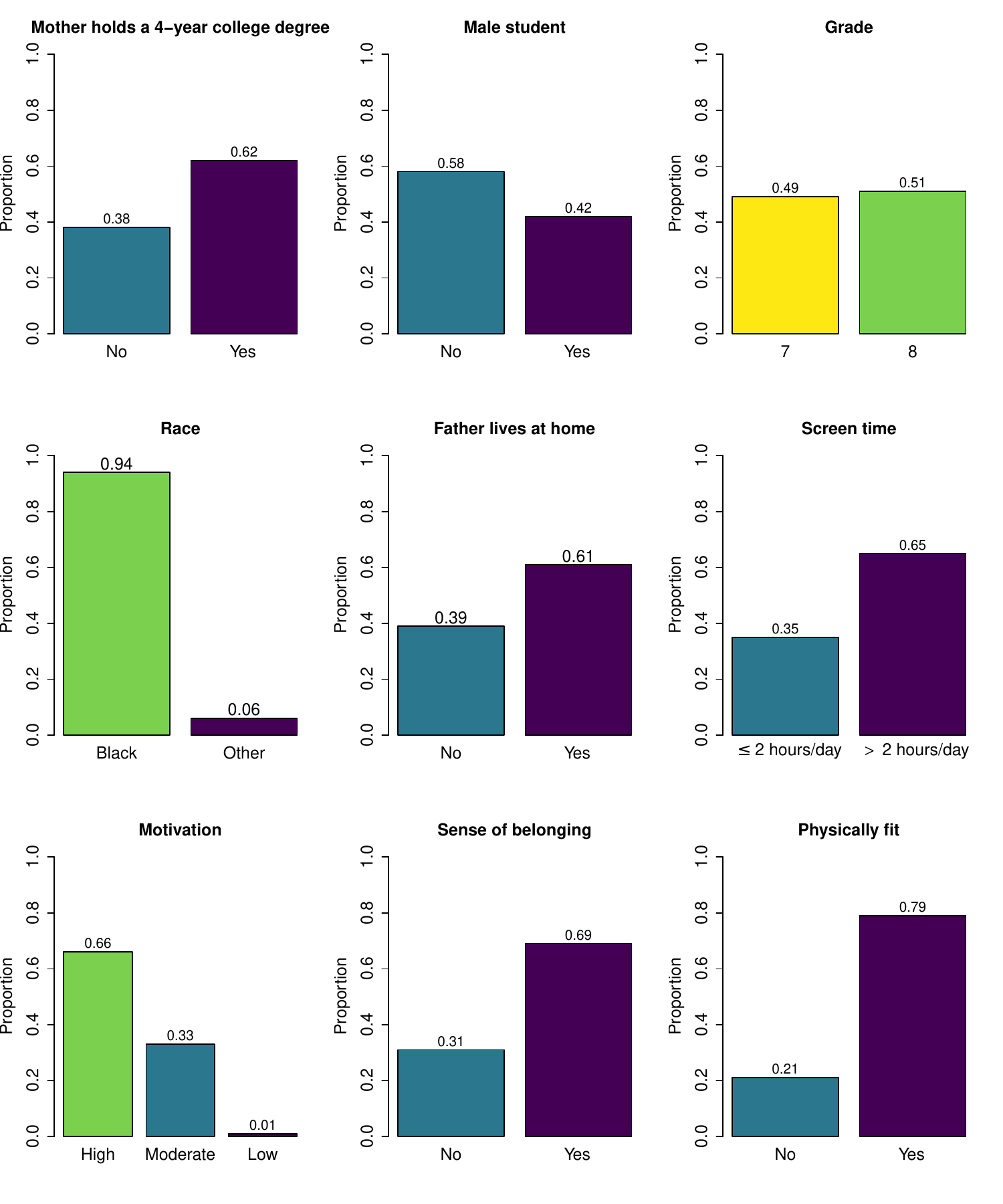}
\caption{Observed marginal distributions for the exposure and selected categorical pretreatment covariates in School 106.}
\label{fig:marginals106}
\end{figure}

\begin{table}[H]
\caption{Estimated Spearman's rank correlation matrix for the exposure and selected categorical attributes in School 106. See \ref{appendix:variables} for variable names.}
\label{tab:spearman106}
\centering
\begin{tabular}{l lllllllll}
  & $Z$ & $X_1$ & $X_2$ & $X_3$ & $X_4$ & $X_5$ & $X_6$ & $X_7$ & $X_8$\\
  \hline
$Z$ & \cellcolor[HTML]{1A1AE6}{ 1.00} & \cellcolor[HTML]{8787F2}{-0.03} & \cellcolor[HTML]{7777F0}{ 0.10} & \cellcolor[HTML]{7777F0}{ 0.07} & \cellcolor[HTML]{8787F2}{ 0.02} & \cellcolor[HTML]{8787F2}{-0.03} & \cellcolor[HTML]{7777F0}{ 0.12} & \cellcolor[HTML]{7777F0}{ 0.10} & \cellcolor[HTML]{8787F2}{ 0.02} \\ 
 $X_1$ & \cellcolor[HTML]{8787F2}{-0.03} & \cellcolor[HTML]{1A1AE6}{ 1.00} & \cellcolor[HTML]{7777F0}{ 0.16} & \cellcolor[HTML]{8787F2}{ 0.04} & \cellcolor[HTML]{8787F2}{ 0.05} & \cellcolor[HTML]{8787F2}{ 0.03} & \cellcolor[HTML]{7777F0}{ 0.18} & \cellcolor[HTML]{8787F2}{ 0.02} & \cellcolor[HTML]{6868EE}{ 0.23} \\ 
 $X_2$ & \cellcolor[HTML]{7777F0}{ 0.10} & \cellcolor[HTML]{7777F0}{ 0.16} & \cellcolor[HTML]{1A1AE6}{ 1.00} & \cellcolor[HTML]{9797F4}{-0.13} & \cellcolor[HTML]{9797F4}{-0.16} & \cellcolor[HTML]{8787F2}{ 0.04} & \cellcolor[HTML]{8787F2}{-0.06} & \cellcolor[HTML]{7777F0}{ 0.13} & \cellcolor[HTML]{7777F0}{ 0.11} \\ 
 $X_3$ & \cellcolor[HTML]{7777F0}{ 0.07} & \cellcolor[HTML]{8787F2}{ 0.04} & \cellcolor[HTML]{9797F4}{-0.13} & \cellcolor[HTML]{1A1AE6}{ 1.00} & \cellcolor[HTML]{8787F2}{-0.05} & \cellcolor[HTML]{7777F0}{ 0.18} & \cellcolor[HTML]{8787F2}{-0.05} & \cellcolor[HTML]{A6A6F6}{-0.23} & \cellcolor[HTML]{9797F4}{-0.17} \\ 
  $X_4$ &  \cellcolor[HTML]{8787F2}{ 0.02} & \cellcolor[HTML]{8787F2}{ 0.05} & \cellcolor[HTML]{9797F4}{-0.16} & \cellcolor[HTML]{8787F2}{-0.05} & \cellcolor[HTML]{1A1AE6}{ 1.00} & \cellcolor[HTML]{A6A6F6}{-0.23} & \cellcolor[HTML]{7777F0}{ 0.16} & \cellcolor[HTML]{6868EE}{ 0.21} & \cellcolor[HTML]{7777F0}{ 0.15} \\ 
 $X_5$ & \cellcolor[HTML]{8787F2}{-0.03} & \cellcolor[HTML]{8787F2}{ 0.03} & \cellcolor[HTML]{8787F2}{ 0.04} & \cellcolor[HTML]{7777F0}{ 0.18} & \cellcolor[HTML]{A6A6F6}{-0.23} & \cellcolor[HTML]{1A1AE6}{ 1.00} & \cellcolor[HTML]{9797F4}{-0.09} & \cellcolor[HTML]{A6A6F6}{-0.30} & \cellcolor[HTML]{9797F4}{-0.09} \\ 
 $X_6$ &  \cellcolor[HTML]{7777F0}{ 0.12} & \cellcolor[HTML]{7777F0}{ 0.18} & \cellcolor[HTML]{8787F2}{-0.06} & \cellcolor[HTML]{8787F2}{-0.05} & \cellcolor[HTML]{7777F0}{ 0.16} & \cellcolor[HTML]{9797F4}{-0.09} & \cellcolor[HTML]{1A1AE6}{ 1.00} & \cellcolor[HTML]{6868EE}{ 0.29} & \cellcolor[HTML]{8787F2}{ 0.01} \\ 
 $X_7$ & \cellcolor[HTML]{7777F0}{ 0.10} & \cellcolor[HTML]{8787F2}{ 0.02} & \cellcolor[HTML]{7777F0}{ 0.13} & \cellcolor[HTML]{A6A6F6}{-0.23} & \cellcolor[HTML]{6868EE}{ 0.21} & \cellcolor[HTML]{A6A6F6}{-0.30} & \cellcolor[HTML]{6868EE}{ 0.29} & \cellcolor[HTML]{1A1AE6}{ 1.00} & \cellcolor[HTML]{7777F0}{ 0.18} \\ 
  $X_8$ & \cellcolor[HTML]{8787F2}{ 0.02} & \cellcolor[HTML]{6868EE}{ 0.23} & \cellcolor[HTML]{7777F0}{ 0.11} & \cellcolor[HTML]{9797F4}{-0.17} & \cellcolor[HTML]{7777F0}{ 0.15} & \cellcolor[HTML]{9797F4}{-0.09} & \cellcolor[HTML]{8787F2}{ 0.01} & \cellcolor[HTML]{7777F0}{ 0.18} & \cellcolor[HTML]{1A1AE6}{ 1.00} \\
   \hline
\end{tabular}
\end{table}

\begin{table}[H]
\caption{Coefficient estimates for the ERGM fitted to School 106}
\label{tab:ergm028}
\centering

\begin{tabular}{lc}
  \toprule
 Term   &\multicolumn{1}{c}{Coefficient estimate} \\
  \midrule
Edges & -5.7537 \\
GWESP ($\lambda = 1$) & 0.6264 \\
GWD ($\lambda = 1$) & 1.3862 \\
NF Male & 0.0478  \\
NF Race (Other) & 0.7784 \\
NF Grade (8) & -0.0003  \\
UH Male & 0.4298 \\
UH Race & 0.7199\\
UH Grade & 1.3721 \\
\bottomrule
\end{tabular}
\scriptsize
\begin{tablenotes}
    \item[1] Abbreviations: GWESP = Geometrically Weighted Edgewise Shared Partner; GWD = Geometrically Weighted Degree; NF = Node Factor; UH = Uniform Homophily.
    \item[2] Baseline levels for NF Race and NF Grade are Black and 7, respectively.
  \end{tablenotes}
\end{table}

\begin{figure}[H]
    \includegraphics[scale=0.45]{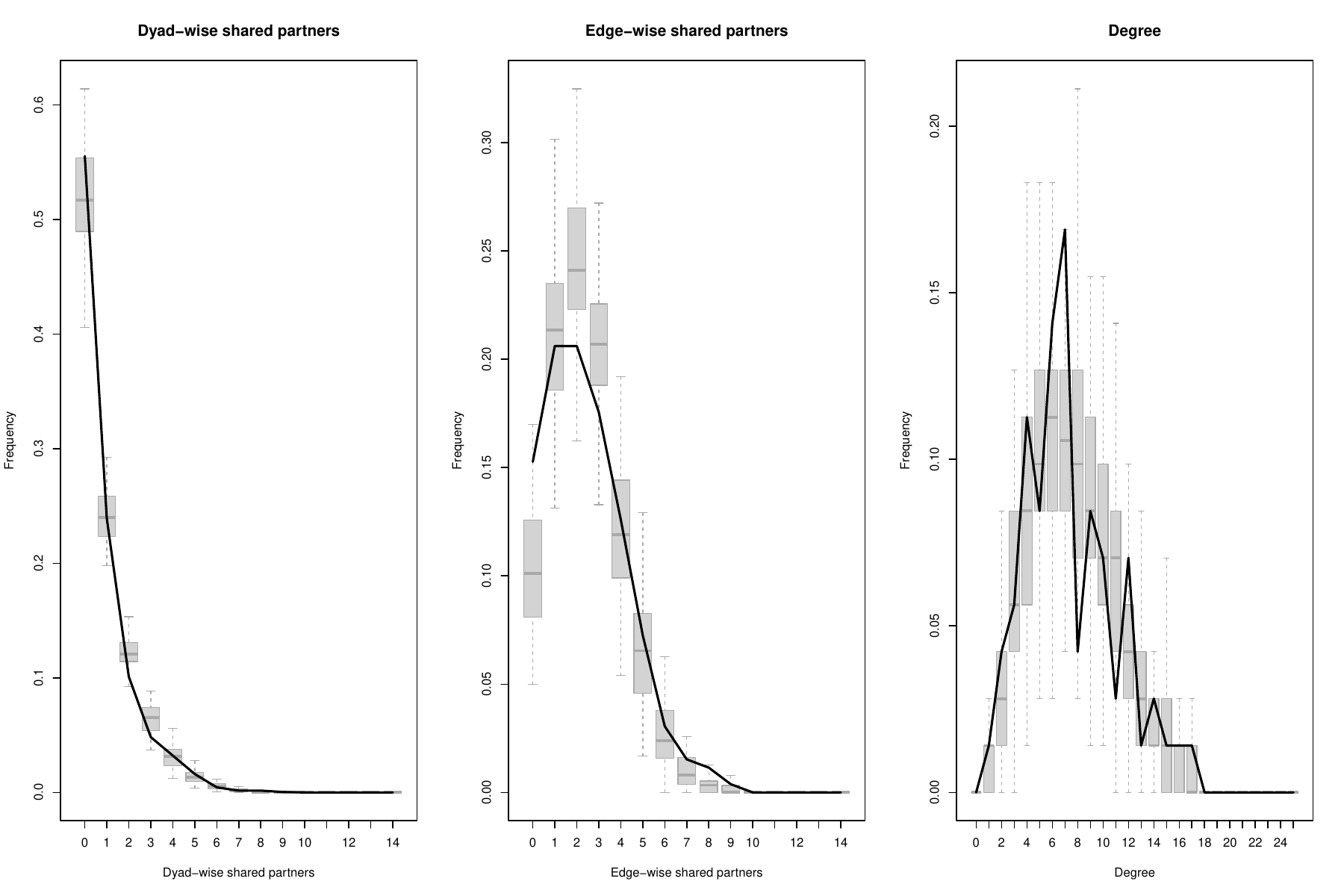}
\caption{Goodness-of-fit assessment of the ERGM for School 106 based on 100 simulated networks. The black solid line represents the observed distribution, whereas the boxplots display the distributions of the network statistics obtained across the 100 simulated networks.}
\label{fig:gof106}
\end{figure}

\section{Sufficient statistics and model estimates for School 122}
\label{appendix:school122}
\begin{figure}[H]
\includegraphics[scale=0.54]{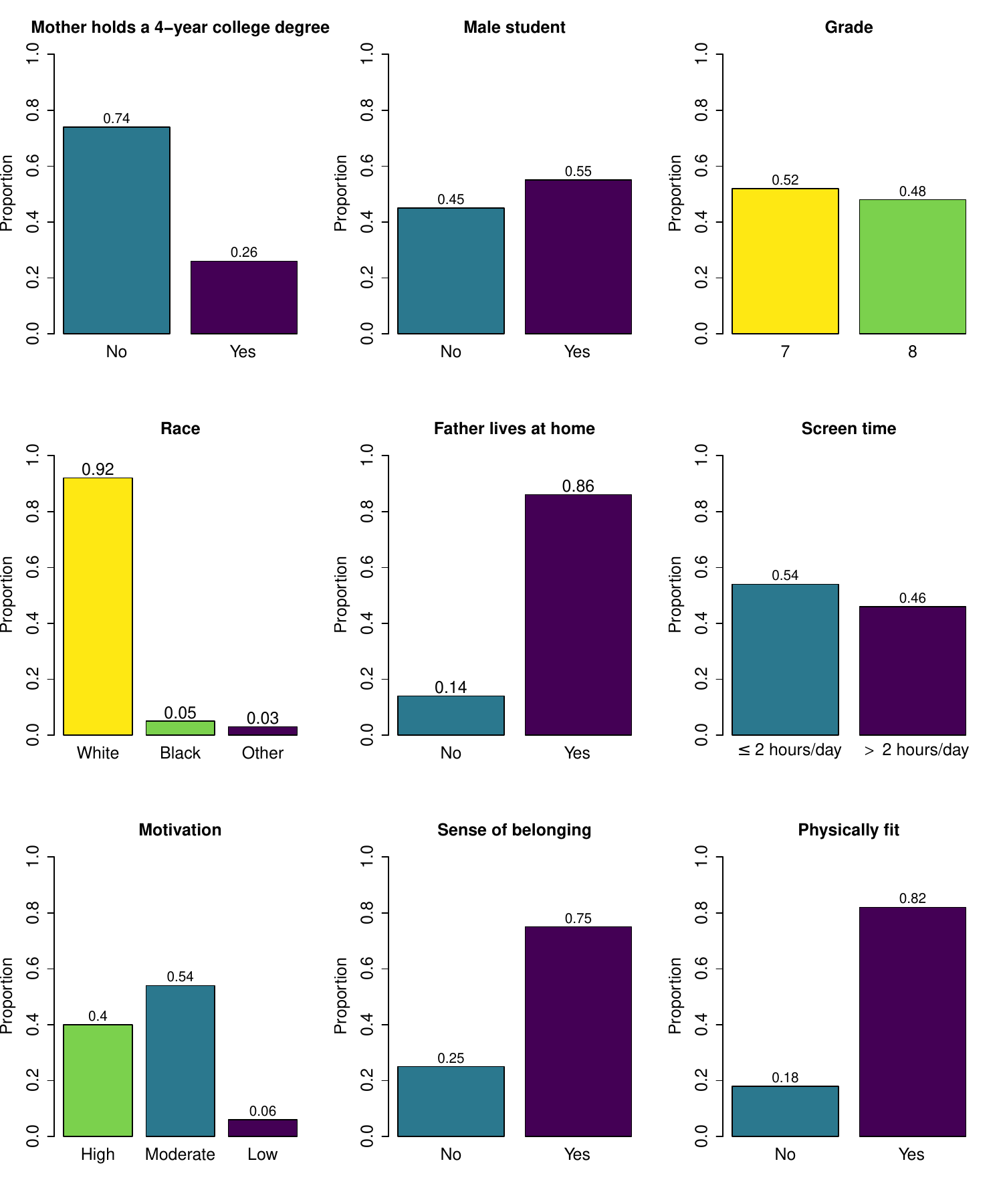}
\caption{Observed marginal distributions for the exposure and selected categorical pretreatment covariates in School 122.}
\label{fig:marginals122}
\end{figure}

\begin{table}[H]
\caption{Estimated Spearman's rank correlation matrix for the exposure and selected categorical attributes in School 122. See \ref{appendix:variables} for variable names.}
\label{tab:spearman122}
\centering
\begin{tabular}{l lllllllll}
  & $Z$ & $X_1$ & $X_2$ & $X_3$ & $X_4$ & $X_5$ & $X_6$ & $X_7$ & $X_8$\\
  \hline
$Z$ & \cellcolor[HTML]{1A1AE6}{ 1.00} & \cellcolor[HTML]{8787F2}{ 0.03} & \cellcolor[HTML]{7777F0}{ 0.07} & \cellcolor[HTML]{9797F4}{-0.08} & \cellcolor[HTML]{8787F2}{-0.03} & \cellcolor[HTML]{9797F4}{-0.08} & \cellcolor[HTML]{8787F2}{ 0.01} & \cellcolor[HTML]{7777F0}{ 0.13} & \cellcolor[HTML]{7777F0}{ 0.16} \\ 
 $X_1$& \cellcolor[HTML]{8787F2}{ 0.03} & \cellcolor[HTML]{1A1AE6}{ 1.00} & \cellcolor[HTML]{8787F2}{-0.02} & \cellcolor[HTML]{8787F2}{-0.04} & \cellcolor[HTML]{6868EE}{ 0.22} & \cellcolor[HTML]{9797F4}{-0.10} & \cellcolor[HTML]{8787F2}{ 0.05} & \cellcolor[HTML]{8787F2}{-0.01} & \cellcolor[HTML]{8787F2}{ 0.05} \\ 
 $X_2$& \cellcolor[HTML]{7777F0}{ 0.07} & \cellcolor[HTML]{8787F2}{-0.02} & \cellcolor[HTML]{1A1AE6}{ 1.00} & \cellcolor[HTML]{7777F0}{ 0.09} & \cellcolor[HTML]{8787F2}{ 0.04} & \cellcolor[HTML]{9797F4}{-0.19} & \cellcolor[HTML]{8787F2}{ 0.05} & \cellcolor[HTML]{8787F2}{ 0.04} & \cellcolor[HTML]{8787F2}{ 0.03} \\ 
 $X_3$ & \cellcolor[HTML]{9797F4}{-0.08} & \cellcolor[HTML]{8787F2}{-0.04} & \cellcolor[HTML]{7777F0}{ 0.09} & \cellcolor[HTML]{1A1AE6}{ 1.00} & \cellcolor[HTML]{A6A6F6}{-0.31} & \cellcolor[HTML]{9797F4}{-0.13} & \cellcolor[HTML]{9797F4}{-0.07} & \cellcolor[HTML]{9797F4}{-0.09} & \cellcolor[HTML]{8787F2}{ 0.04} \\ 
 $X_4$ &\cellcolor[HTML]{8787F2}{-0.03} & \cellcolor[HTML]{6868EE}{ 0.22} & \cellcolor[HTML]{8787F2}{ 0.04} & \cellcolor[HTML]{A6A6F6}{-0.31} & \cellcolor[HTML]{1A1AE6}{ 1.00} & \cellcolor[HTML]{7777F0}{ 0.14} & \cellcolor[HTML]{8787F2}{ 0.02} & \cellcolor[HTML]{7777F0}{ 0.10} & \cellcolor[HTML]{8787F2}{ 0.04} \\ 
 $X_5$ & \cellcolor[HTML]{9797F4}{-0.08} & \cellcolor[HTML]{9797F4}{-0.10} & \cellcolor[HTML]{9797F4}{-0.19} & \cellcolor[HTML]{9797F4}{-0.13} & \cellcolor[HTML]{7777F0}{ 0.14} & \cellcolor[HTML]{1A1AE6}{ 1.00} & \cellcolor[HTML]{7777F0}{ 0.07} & \cellcolor[HTML]{8787F2}{-0.03} & \cellcolor[HTML]{A6A6F6}{-0.25} \\ 
 $X_6$ & \cellcolor[HTML]{8787F2}{ 0.01} & \cellcolor[HTML]{8787F2}{ 0.05} & \cellcolor[HTML]{8787F2}{ 0.05} & \cellcolor[HTML]{9797F4}{-0.07} & \cellcolor[HTML]{8787F2}{ 0.02} & \cellcolor[HTML]{7777F0}{ 0.07} & \cellcolor[HTML]{1A1AE6}{ 1.00} & \cellcolor[HTML]{A6A6F6}{-0.32} & \cellcolor[HTML]{8787F2}{ 0.01} \\ 
 $X_7$ & \cellcolor[HTML]{7777F0}{ 0.13} & \cellcolor[HTML]{8787F2}{-0.01} & \cellcolor[HTML]{8787F2}{ 0.04} & \cellcolor[HTML]{9797F4}{-0.09} & \cellcolor[HTML]{7777F0}{ 0.10} & \cellcolor[HTML]{8787F2}{-0.03} & \cellcolor[HTML]{A6A6F6}{-0.32} & \cellcolor[HTML]{1A1AE6}{ 1.00} & \cellcolor[HTML]{6868EE}{ 0.27} \\ 
 $X_8$ & \cellcolor[HTML]{7777F0}{ 0.16} & \cellcolor[HTML]{8787F2}{ 0.05} & \cellcolor[HTML]{8787F2}{ 0.03} & \cellcolor[HTML]{8787F2}{ 0.04} & \cellcolor[HTML]{8787F2}{ 0.04} & \cellcolor[HTML]{A6A6F6}{-0.25} & \cellcolor[HTML]{8787F2}{ 0.01} & \cellcolor[HTML]{6868EE}{ 0.27} & \cellcolor[HTML]{1A1AE6}{ 1.00} \\ 
   \hline
\end{tabular}
\end{table}

\begin{table}[H]
\caption{Coefficient estimates for the ERGM fitted to School 122}
\label{tab:ergm028}
\centering

\begin{tabular}{lc}
  \toprule
 Term   &\multicolumn{1}{c}{Coefficient estimate} \\
  \midrule
Edges & -5.2088 \\
GWESP ($\lambda = 1$) & 1.0589 \\
GWD ($\lambda = 1$) & 2.7566 \\
NF Male & -0.0608  \\
NF Race (Other) & -0.3427 \\
NF Race (White) & -0.8059 \\
NF Grade (8) & 0.0249  \\
UH Male & 0.3217 \\
UH Race & 0.8190\\
UH Grade & 0.9421 \\
\bottomrule
\end{tabular}
\scriptsize
\begin{tablenotes}
    \item[1] Abbreviations: GWESP = Geometrically Weighted Edgewise Shared Partner; GWD = Geometrically Weighted Degree; NF = Node Factor; UH = Uniform Homophily. \item[2] Baseline levels for NF Race and NF Grade are Black and 7, respectively.
  \end{tablenotes}
\end{table}

\begin{figure}[H]
    \includegraphics[scale=0.45]{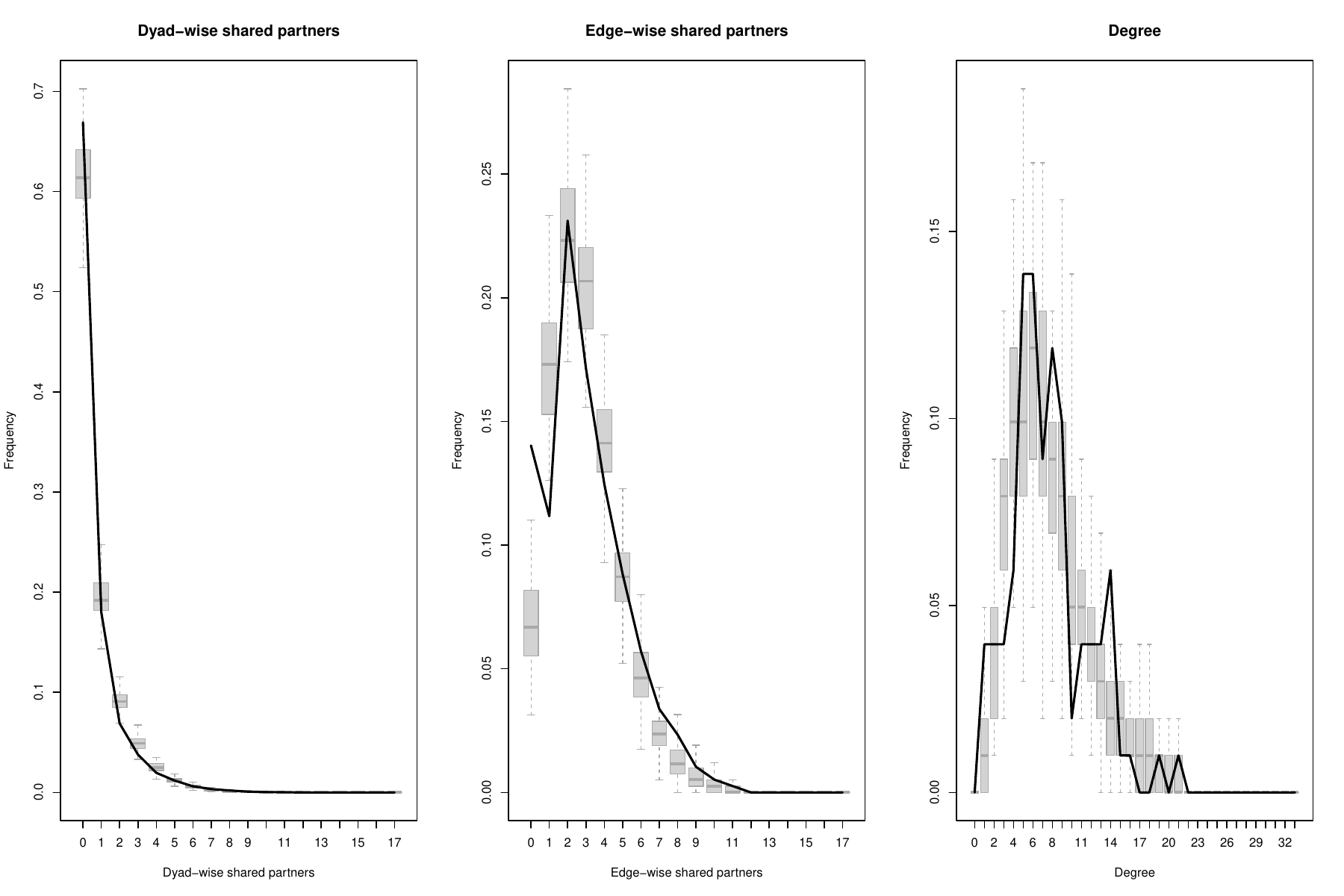}
    \caption{Goodness-of-fit assessment of the ERGM for School 122 based on 100 simulated networks. The black solid line represents the observed distribution, whereas the boxplots display the distributions of the network statistics obtained across the 100 simulated networks.}
\label{fig:gof122}
\end{figure}

\section{Sufficient statistics and model estimates for School 173}
\label{appendix:school173}
\begin{figure}[H]
\includegraphics[scale=0.54]{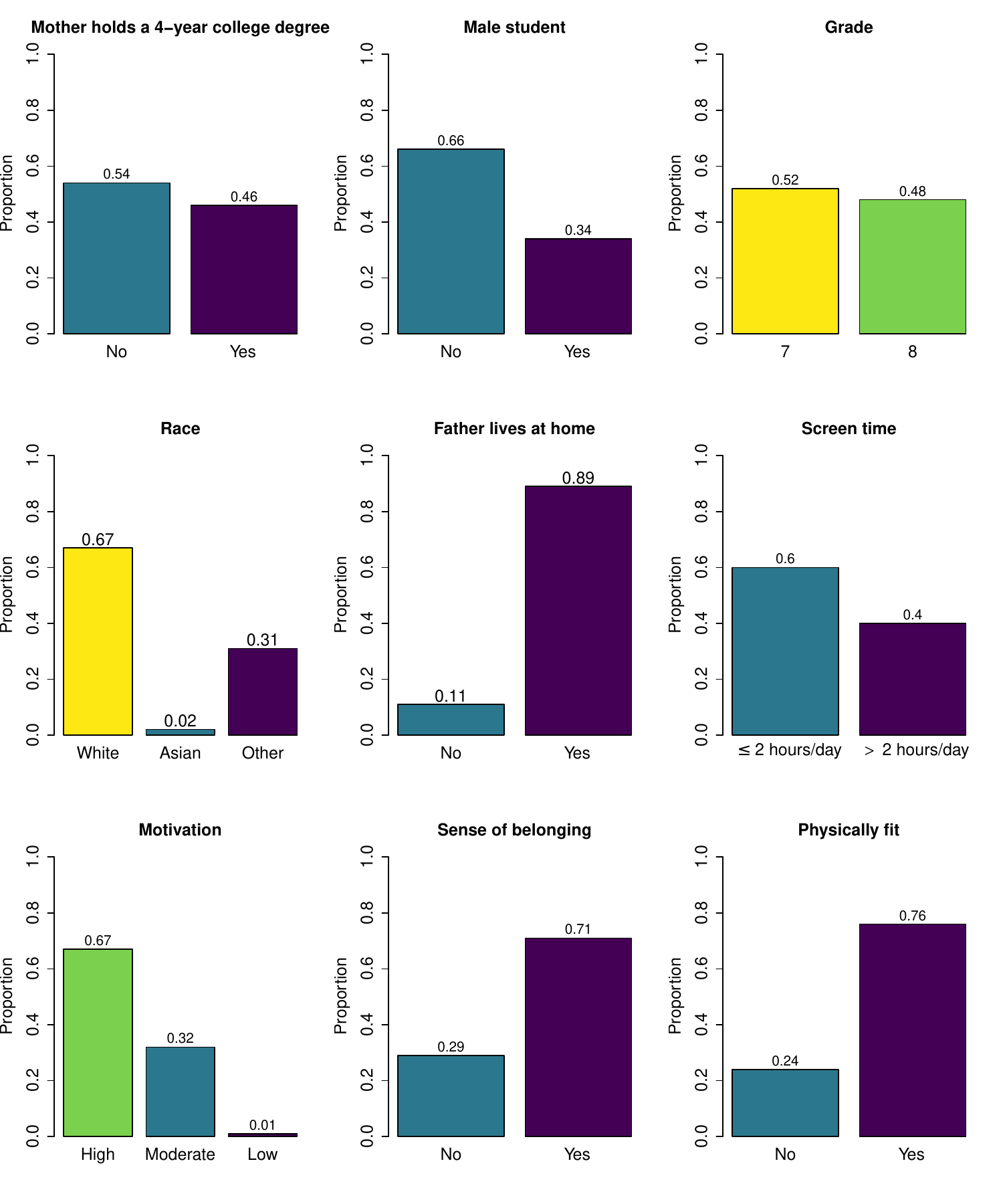}
\caption{Observed marginal distributions for the exposure and selected categorical pretreatment covariates in School 173.}
\label{fig:marginals173}
\end{figure}

\begin{table}[H]
\caption{Estimated Spearman's rank correlation matrix for the exposure and selected categorical attributes in School 173. See \ref{appendix:variables} for variable names.}
\label{tab:spearman173}
\centering
\begin{tabular}{l lllllllll}
  & $Z$ & $X_1$ & $X_2$ & $X_3$ & $X_4$ & $X_5$ & $X_6$ & $X_7$ & $X_8$\\
  \hline
$Z$ & \cellcolor[HTML]{1A1AE6}{ 1.00} & \cellcolor[HTML]{6868EE}{ 0.30} & \cellcolor[HTML]{7777F0}{ 0.08} & \cellcolor[HTML]{9797F4}{-0.13} & \cellcolor[HTML]{7777F0}{ 0.12} & \cellcolor[HTML]{9797F4}{-0.15} & \cellcolor[HTML]{9797F4}{-0.10} & \cellcolor[HTML]{9797F4}{-0.08} & \cellcolor[HTML]{7777F0}{ 0.14} \\ 
 $X_1$ & \cellcolor[HTML]{6868EE}{ 0.30} & \cellcolor[HTML]{1A1AE6}{ 1.00} & \cellcolor[HTML]{7777F0}{ 0.17} & \cellcolor[HTML]{9797F4}{-0.11} & \cellcolor[HTML]{7777F0}{ 0.11} & \cellcolor[HTML]{7777F0}{ 0.14} & \cellcolor[HTML]{6868EE}{ 0.31} & \cellcolor[HTML]{A6A6F6}{-0.29} & \cellcolor[HTML]{7777F0}{ 0.13} \\ 
 $X_2$ & \cellcolor[HTML]{7777F0}{ 0.08} & \cellcolor[HTML]{7777F0}{ 0.17} & \cellcolor[HTML]{1A1AE6}{ 1.00} & \cellcolor[HTML]{8787F2}{-0.03} & \cellcolor[HTML]{6868EE}{ 0.20} & \cellcolor[HTML]{9797F4}{-0.14} & \cellcolor[HTML]{7777F0}{ 0.09} & \cellcolor[HTML]{8787F2}{ 0.00} & \cellcolor[HTML]{6868EE}{ 0.28} \\ 
  $X_3$ & \cellcolor[HTML]{9797F4}{-0.13} & \cellcolor[HTML]{9797F4}{-0.11} & \cellcolor[HTML]{8787F2}{-0.03} & \cellcolor[HTML]{1A1AE6}{ 1.00} & \cellcolor[HTML]{A6A6F6}{-0.21} & \cellcolor[HTML]{8787F2}{-0.05} & \cellcolor[HTML]{9797F4}{-0.08} & \cellcolor[HTML]{A6A6F6}{-0.21} & \cellcolor[HTML]{9797F4}{-0.11} \\ 
 $X_4$ & \cellcolor[HTML]{7777F0}{ 0.12} & \cellcolor[HTML]{7777F0}{ 0.11} & \cellcolor[HTML]{6868EE}{ 0.20} & \cellcolor[HTML]{A6A6F6}{-0.21} & \cellcolor[HTML]{1A1AE6}{ 1.00} & \cellcolor[HTML]{8787F2}{ 0.00} & \cellcolor[HTML]{9797F4}{-0.07} & \cellcolor[HTML]{7777F0}{ 0.17} & \cellcolor[HTML]{5858ED}{ 0.39} \\ 
 $X_5$ &  \cellcolor[HTML]{9797F4}{-0.15} & \cellcolor[HTML]{7777F0}{ 0.14} & \cellcolor[HTML]{9797F4}{-0.14} & \cellcolor[HTML]{8787F2}{-0.05} & \cellcolor[HTML]{8787F2}{ 0.00} & \cellcolor[HTML]{1A1AE6}{ 1.00} & \cellcolor[HTML]{5858ED}{ 0.35} & \cellcolor[HTML]{A6A6F6}{-0.31} & \cellcolor[HTML]{A6A6F6}{-0.30} \\ 
 $X_6$ & \cellcolor[HTML]{9797F4}{-0.10} & \cellcolor[HTML]{6868EE}{ 0.31} & \cellcolor[HTML]{7777F0}{ 0.09} & \cellcolor[HTML]{9797F4}{-0.08} & \cellcolor[HTML]{9797F4}{-0.07} & \cellcolor[HTML]{5858ED}{ 0.35} & \cellcolor[HTML]{1A1AE6}{ 1.00} & \cellcolor[HTML]{A6A6F6}{-0.26} & \cellcolor[HTML]{9797F4}{-0.18} \\ 
 $X_7$ & \cellcolor[HTML]{9797F4}{-0.08} & \cellcolor[HTML]{A6A6F6}{-0.29} & \cellcolor[HTML]{8787F2}{ 0.00} & \cellcolor[HTML]{A6A6F6}{-0.21} & \cellcolor[HTML]{7777F0}{ 0.17} & \cellcolor[HTML]{A6A6F6}{-0.31} & \cellcolor[HTML]{A6A6F6}{-0.26} & \cellcolor[HTML]{1A1AE6}{ 1.00} & \cellcolor[HTML]{6868EE}{ 0.24} \\ 
 $X_8$ & \cellcolor[HTML]{7777F0}{ 0.14} & \cellcolor[HTML]{7777F0}{ 0.13} & \cellcolor[HTML]{6868EE}{ 0.28} & \cellcolor[HTML]{9797F4}{-0.11} & \cellcolor[HTML]{5858ED}{ 0.39} & \cellcolor[HTML]{A6A6F6}{-0.30} & \cellcolor[HTML]{9797F4}{-0.18} & \cellcolor[HTML]{6868EE}{ 0.24} & \cellcolor[HTML]{1A1AE6}{ 1.00} \\ 
   \hline
\end{tabular}
\end{table}

\begin{table}[H]
\caption{Coefficient estimates for the ERGM fitted to School 173}
\label{tab:ergm028}
\centering

\begin{tabular}{lc}
  \toprule
 Term   &\multicolumn{1}{c}{Coefficient estimate} \\
  \midrule
Edges & -5.9457 \\
GWESP ($\lambda = 1$) & 1.0954 \\
GWD ($\lambda = 1$) & 1.3935 \\
NF Male & 0.1130  \\
NF Race (Other) & 0.2725 \\
NF Race (White) & 0.1891 \\
NF Grade (8) & -0.1705  \\
UH Male & 0.3130 \\
UH Race & 0.2060\\
UH Grade & -0.0287 \\
\bottomrule
\end{tabular}
\scriptsize
\begin{tablenotes}
    \item[1] Abbreviations: GWESP = Geometrically Weighted Edgewise Shared Partner; GWD = Geometrically Weighted Degree; NF = Node Factor; UH = Uniform Homophily. \item[2] Baseline levels for NF Race and NF Grade are Asian and 7, respectively.
  \end{tablenotes}
\end{table}

\begin{figure}[H]
    \includegraphics[scale=0.45]{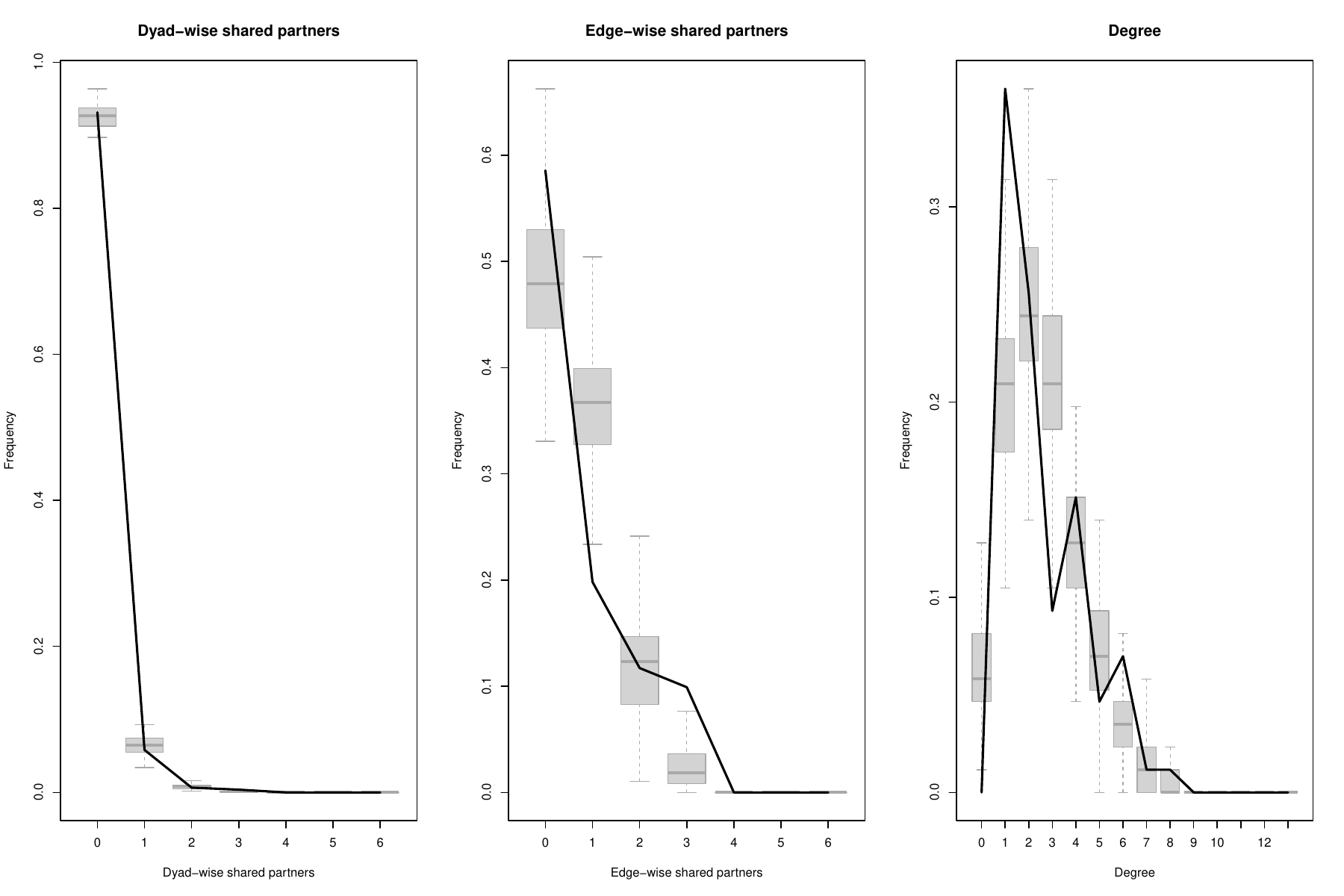}
    \caption{Goodness-of-fit assessment of the ERGM for School 173 based on 100 simulated networks. The black solid line represents the observed distribution, whereas the boxplots display the distributions of the network statistics obtained across the 100 simulated networks.}
\label{fig:gof173}
\end{figure}


\section{Parameter estimates for the outcome model}
\label{appendix:outcomemodel}
\begin{table}[H]
\caption{Estimated linear mixed model for the outcome GPA fitted on the complete eligible study population (139 schools, $N=73,580$)}
\label{tab:outcomemodel}
\centering

\begin{tabular}{lc}
  \toprule
 Parameter\textsuperscript{a}   &\multicolumn{1}{c}{Estimate} \\
  \midrule
Intercept (${\beta}_0$) & 0.001233 \\
Maternal education ($\beta_Z$) & 0.321031 \\
Neighborhood treatment ($\beta_{\bm{Z}_{\mathcal{N}}} $) & 0.412686 \\
Maternal education $\times$ Neighborhood treatment ($\beta_{Z \times \bm{Z}_{\mathcal{N}}}$)  & -0.077339  \\
Grade (8) ($\beta_1$) & 0.010047  \\
Grade (9) ($\beta_2$) & -0.032272 \\
Grade (10) ($\beta_3$) & 0.003732 \\
Grade (11) ($\beta_4$) & 0.032506  \\
Grade (12) ($\beta_5$) & 0.134479 \\
Male ($\beta_6$) & -0.176407 \\
Race (Black) ($\beta_7$)  & -0.306032 \\
Race (Asian) ($\beta_8$)  & 0.241087 \\
Race (Other) ($\beta_9$) & -0.166038 \\
Father lives at home ($\beta_{10}$) & 0.160135\\
Screen time ($> $ 2 hours/day) ($\beta_{11}$) & -0.119603\\
Motivation (Moderate) ($\beta_{12}$) & -0.342066\\
Motivation (Low) ($\beta_{13}$) & -0.737789\\
Sense of belonging ($\beta_{14}$) & 0.118153\\
Physically fit ($\beta_{15}$) & 0.122159\\
\midrule
Residual variance ($\sigma^2_{\varepsilon})$ & 0.76476\\
Between-school variance ($\sigma^2_{b_Y}$) & 0.03822\\ 
\bottomrule
\end{tabular}
\centering
\begin{tablenotes}
    \item[2]\textsuperscript{a} Baseline levels for Grade, Race, and Motivation are 7, White, and High, respectively.
  \end{tablenotes}
\end{table}

\section{Parameter estimates for the exposure model}
\label{appendix:exposuremodel}

\begin{table}[H]
\caption{Estimated mixed effects logistic model for the exposure $Z$ fitted on the complete eligible study population (139 schools, $N=73,580$)}
\label{tab:exposuremodel}
\centering

\begin{tabular}{lc}
  \toprule
 Parameter\textsuperscript{a}   &\multicolumn{1}{c}{Estimate} \\
  \midrule
Intercept (${\gamma}_0$) & -0.923252 \\
Race (Black) ($\gamma_1$)  & -0.174938 \\
Race (Asian) ($\gamma_2$)  & 0.533800 \\
Race (Other) ($\gamma_3$) & -0.520414 \\
Father lives at home ($\gamma_{4}$) & 0.208974\\
\midrule
Between-school variance ($\sigma^2_{b_Z}$) & 0.395118\\ 
\bottomrule
\end{tabular}
\centering
\begin{tablenotes}
    \item[2]\textsuperscript{a} The baseline level for Race is White.
  \end{tablenotes}
\end{table}





\end{document}